\documentclass[fleqn,usenatbib]{aa}
\usepackage{txfonts}
\usepackage{graphicx}
\usepackage{natbib}
\usepackage{longtable}
\usepackage{subeqnarray}
\usepackage{cases}
\usepackage[colorlinks=true,citecolor=blue]{hyperref}
\usepackage{amsmath}
\newcommand*{\mk}{ }
\usepackage[switch]{lineno} 

\newcommand{\orcid}[1]{\href{https://orcid.org/#1}{\includegraphics[width=10pt]{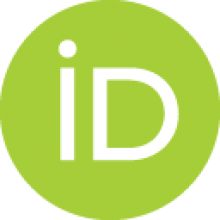}}}

\begin{document}
\title{The Impact of Expanding HII Regions on Filament G37}
\subtitle{Curved Magnetic Field and Multiple Direction Material Flows}
   
\author{Mengke Zhao\orcid{[0000-0003-0596-6608]}\inst{\ref{inst1},\ref{inst2},\ref{inst3}}
\and Xindi Tang\orcid{0000-0002-4154-4309}\inst{\ref{inst2},\ref{inst3},\ref{inst4},\ref{inst5}}
\and Keping Qiu\orcid{0000-0002-5093-5088}\inst{\ref{inst1},\ref{inst6}}
\and Yuxin He\orcid{0000-0002-8760-8988}\inst{\ref{inst2},\ref{inst3},\ref{inst4},\ref{inst5}}
\and Dalei Li\inst{\ref{inst2},\ref{inst3},\ref{inst4},\ref{inst5}}
}

\titlerunning{The Impact of Filament G37 on Expanding H\,{\scriptsize II} Regions}
\authorrunning{Zhao et al.}

\institute{
School of Astronomy and Space Science, Nanjing University, 163 Xianlin Avenue, Nanjing 210023, PR China \label{inst1} \\
\email{mkzhao628@gmail.com} 
\and Xinjiang Astronomical Observatory, Chinese Academy of Sciences, Urumqi 830011, PR China \label{inst2} \\ 
\email{tangxindi@xao.ac.cn} 
\and University of Chinese Academy of Sciences, Beijing 100049, PR China \label{inst3} 
\and Key Laboratory of Radio Astronomy and Technology (Chinese Academy of Sciences), A20 Datun Road, Chaoyang District, Beijing 100101, PR China \label{inst4} 
\and Xinjiang Key Laboratory of Radio Astrophysics, Urumqi 830011, PR China \label{inst5}
\and Key Laboratory of Modern Astronomy and Astrophysics (Nanjing University), Ministry of Education, Nanjing 210023, Jiangsu, PR China \label{inst6}
} 

\abstract
{Filament G37 exhibits a distinctive "caterpillar" shape, characterized by two semicircular structures within its 40\,pc-long body, 
providing an ideal target to investigate the formation and evolution of filaments. By analyzing multiple observational data, such as 
CO spectral line, the H$\alpha$\,RRL, and multi-wavelength continuum, we find that the expanding H\,{\scriptsize II} regions
surrounding filament G37 exert pressure on the structure of the filament body, which kinetic process present as the gas flows in 
multiple directions along its skeleton. The curved magnetic field structure of filament G37 derived by employing the Velocity 
Gradient Technique with CO is found to be parallel to the filament body and keeps against the pressure from expanded 
H\,{\scriptsize II} regions. The multi-directional flows in the 
filament G37 could cause the accumulation and subsequent collapse of gas, resulting in the formation of massive clumps. 
The curved structure and star formation observed in filament G37 are likely a result of the filament body being squeezed 
by the expanding H\,{\scriptsize II} region.
This physical process occurs over a timescale of approximately 5\,Myr. The filament G37 provides a potential candidate for 
end-dominated collapse.}


\maketitle

\section{Introduction}
In the Milky Way, giant molecular clouds represent the coldest and densest regions \citep{1969ApJ...155L.149F,1977ApJ...218..148M}. 
These regions exhibit intricate filamentary structures that span a wide range of environments within the interstellar medium (ISM)
(e.g., \citealt{1979ApJS...41...87S,1987ApJ...312L..45B,2000prpl.conf...97W,2008A&A...487..993K,2010A&A...518L.103M,2018ApJ...864..153Z}). 
These filamentary structures are believed to play a 
crucial role in the process of star formation \citep{2014prpl.conf...27A}. Notably, the analysis of {\it Herschel} data has revealed a 
strong correlation between filamentary clouds and the occurrence of star formation 
\citep{2010A&A...518L.102A,2014prpl.conf...27A,2015MNRAS.450.4043W,2018MNRAS.473.4890S}. Specifically, filament structures of 
various sizes within the ISM have been closely associated with star formation 
(e.g., \citealt{1979ApJS...41...87S, 1987ApJ...312L..45B,2013A&A...554A..55H,2013A&A...559A..34L,2014ApJS..212....1A,2017A&A...602L...2H,2019ApJ...877....1D,2020A&A...637A..67Y,2020ApJ...899..167B,2022ApJ...930..169B}). Therefore, comprehending the physical origins and evolutionary processes of these filaments is crucial for a comprehensive 
understanding of star formation as a whole \citep{2023ASPC..534..153H}.

\begin{figure*}
    \centering
    \includegraphics[width = 18.5cm]{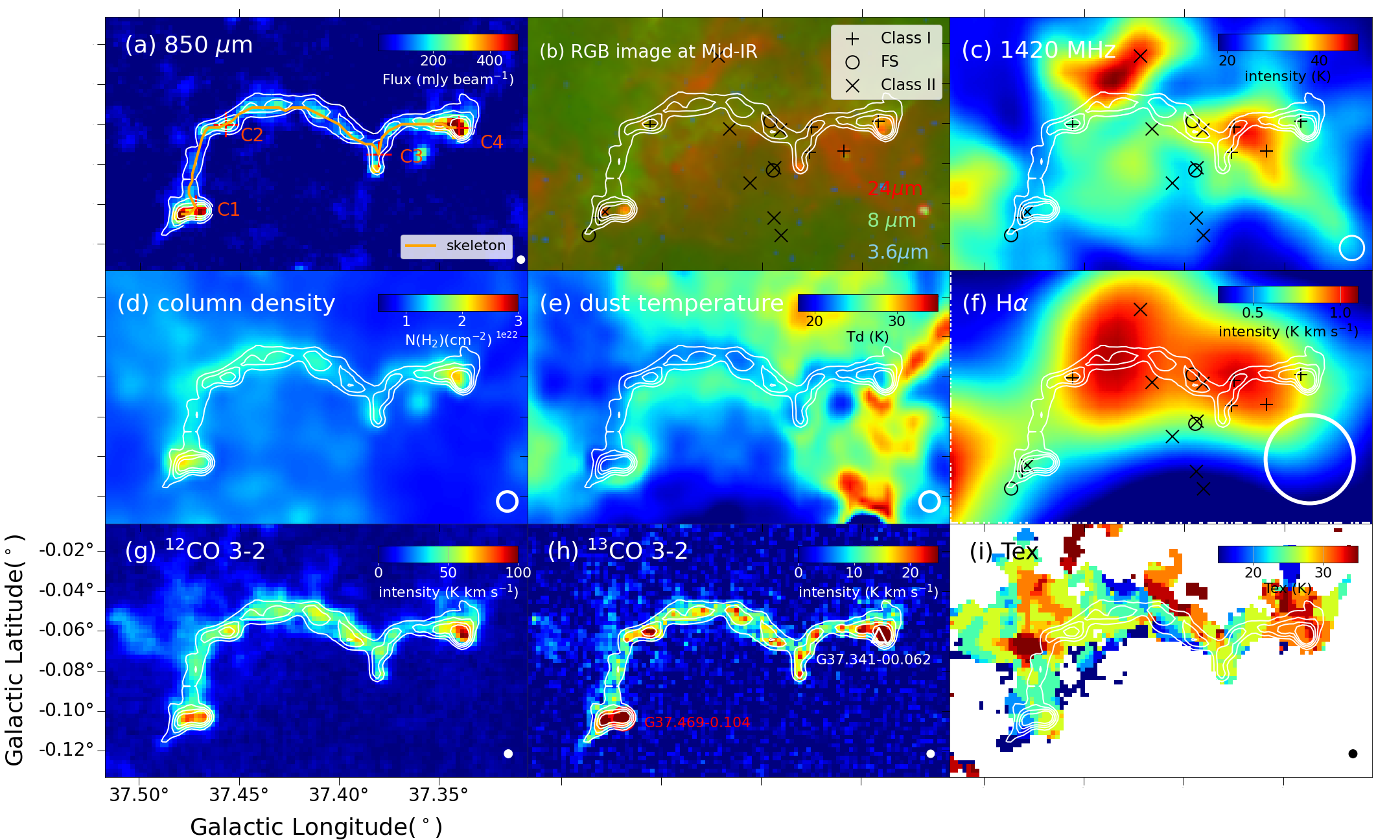}
    \caption{Structure of filament G37. 
    Panels (a), (b), and (c) display the structure of filament G37 at different wavelengths.
    Panel (a) presents the sub-mm wavelength emission at 850\,$\mu$m \citep{2017MNRAS.469.2163E}.
    The orange line displays the skeleton of filament G37, which is identified in \cite{2016A&A...591A...5L}. 
    Four special dense clumps (C1--C4) are denoted by red crosses (see Sect.\,\ref{Overview}).
    The RGB image at the mid-infrared ({\it Spitzer}, GLIMPSE \citealt{2003PASP..115..953B}; MIPSGAL \citealt{2009PASP..121...76C}) 
    is shown in panel (b), where red represents 24\,$\mu$m, green represents 8\,$\mu$m, and blue represents 3.6\,$\mu$m. 
    Panel (c) displays the distribution of continuum emission at 1420\,MHz from VGPS \citep{2006AJ....132.1158S}. 
    The cross markers in panels (b) and (c) display the position of the young stellar objects (YSOs; \citealt{2021ApJS..254...33K}), which uses multiwavelength color and magnitude distributions derived from the mid-infrared range of 3--9\,$\mu$m obtained from the {\it Spitzer} telescope.
    Panels (d) and (e) display the distribution of H$_2$ column density and dust temperature \citep{2018ApJ...864..153Z}. 
    The integrated intensity maps of H$\alpha$\,RRL, $^{12}$CO\,(3-2), and $^{13}$CO\,(3-2) emissions from velocity 51 to 62\,km\,s$^{-1}$ 
    are shown in the panels (f), (g), and (h), respectively. The white contours in all panels show the integrated intensity of $^{13}$CO\,(3--2) 
    from 3 to 15\,K\,km$^{-1}$ with steps of 3\,K\,km$^{-1}$. The distribution of excitation temperature derived
    from $^{12}$CO\,(3--2) emission is shown in panel (i). The spatial resolution of each panel is shown in the lower right corner.}
\label{fig1}
\end{figure*}

The formation and evolution of filaments within the ISM pose significant unresolved questions 
(e.g., \citealt{2013A&A...559A..34L,2016A&A...591A...5L,2018ApJ...864..153Z,2023ASPC..534..153H}). Understanding the kinematics of the ISM, 
including gas velocity information, is crucial for elucidating the radial velocities of gas filaments and their relationship to 
star formation processes \citep{2016A&A...590A...2S,2018A&A...610A..77H}. By examining gas radial velocities, valuable insights 
can be gained into the physical properties and internal dynamics of observed filaments 
\citep{2011A&A...533A..34H,2013A&A...554L...2Z,2017A&A...602L...2H,2018A&A...610A..77H,2018ApJ...855....9L}. 
The kinematics of gas play a pivotal role in unraveling the underlying physical mechanisms involved in filament formation 
and the conversion of gas mass into stellar mass within filaments 
(e.g., \citealt{2014A&A...567A..10L,2019MNRAS.487.1259L,2020A&A...637A..67Y,2023ApJ...957...61H,2023A&A...676A..15M}). 
Filaments in Milky Way with multiple various morphology \citep{2015MNRAS.450.4043W,2016ApJS..226....9W} is a not clearly understood problem which shape of molecular clouds could be caused by squeezed by nearby HII region \citep{2013A&A...559A..34L,2022ARA&A..60..247W,2022Natur.601..334Z}.
Furthermore, the impact of star formation activities in the ISM is noteworthy 
(e.g., \citealt{2016A&A...585A.117Z,2021A&A...647A..78A,2022A&A...660A..56A}). These activities, such as outflow, radiation, 
shocks, and stellar wind, are common phenomena that impact and reshape the structure of the ISM 
(e.g., \citealt{2006ApJ...649..759C,2007ApJ...670..428C,2014MNRAS.438..426H}). 
Additionally, magnetic field plays an important role in star formation and filament evolution but the detail is not understood clearly
(e.g., \citealt{2012ARA&A..50...29C,2019FrASS...6....3H,2021A&A...647A..78A,2021ApJ...923L...9D,2021Galax...9...41L,2022ApJ...941..122C,2022ApJ...941...51H,2022ApJ...926..163K,2024ApJ...962..136W,2024ApJ...976..209Z}). 
Regions of active star formation exhibit similarities in the morphology of gas and magnetic fields 
 \citep{2017ApJ...846..122P,2020ApJ...899...28D,2021A&A...647A..78A}.

The filament G037.410-0.070 (after here, G37) located at $l$\,=\,37.4$^\circ$ 
and $b$\,=\,-0.03$^\circ$ exhibits a distinctive spectral shape resembling a "caterpillar" when observed at sub-millimeter
and far-infrared wavelengths \citep{2016A&A...591A...5L,2016MNRAS.456.2885R,2018ApJ...864..152Z}. In near- and mid-infrared images, 
three bright objects can be identified within the filament body \citep{2016ApJS..226....9W}. This filament was identified by 
\cite{2016A&A...591A...5L}, \cite{2016MNRAS.456.2885R}, \cite{2016ApJS..226....9W}, 
and \cite{2018ApJ...864..152Z}, presents an ideal target to investigate the impact of dynamics, H\,{\scriptsize II} region expansion, 
and magnetic fields on filament and star formation. Extensive data from various surveys have covered this region, including CO observations
\citep{2006ApJS..163..145J,Dempsey2013,2016ApJ...823...77R,2016MNRAS.456.2885R,2017PASJ...69...78U}, 
infrared continuum measurements \citep{2003PASP..115..953B,2009PASP..121...76C,2010A&A...518L.103M,2010A&A...518L...2P,2010A&A...518L...3G}, 
and centimeter wavelength observations \citep{2006AJ....132.1158S}. The unique morphology of filament G37 provides a valuable sample for
studying filament formation and evolution.

In this study, our objective is to investigate the dynamics and magnetic field properties of filament G37, as well as explore the potential 
mechanisms involved in filament formation. The paper is structured as follows. The archival data used in this study are detailed in
Sect.\,\ref{sect2}. The basic physical, dynamics, and magnetic field properties of filament G37 are presented in Sect.\,\ref{sect3}. 
The potential formation mechanisms of filament G37 are explored in Sect.\,\ref{sect4}. A summary of the findings is provided in Sect.\,\ref{sect5}.

\begin{figure*}
\centering
\includegraphics[width = 18.5cm]{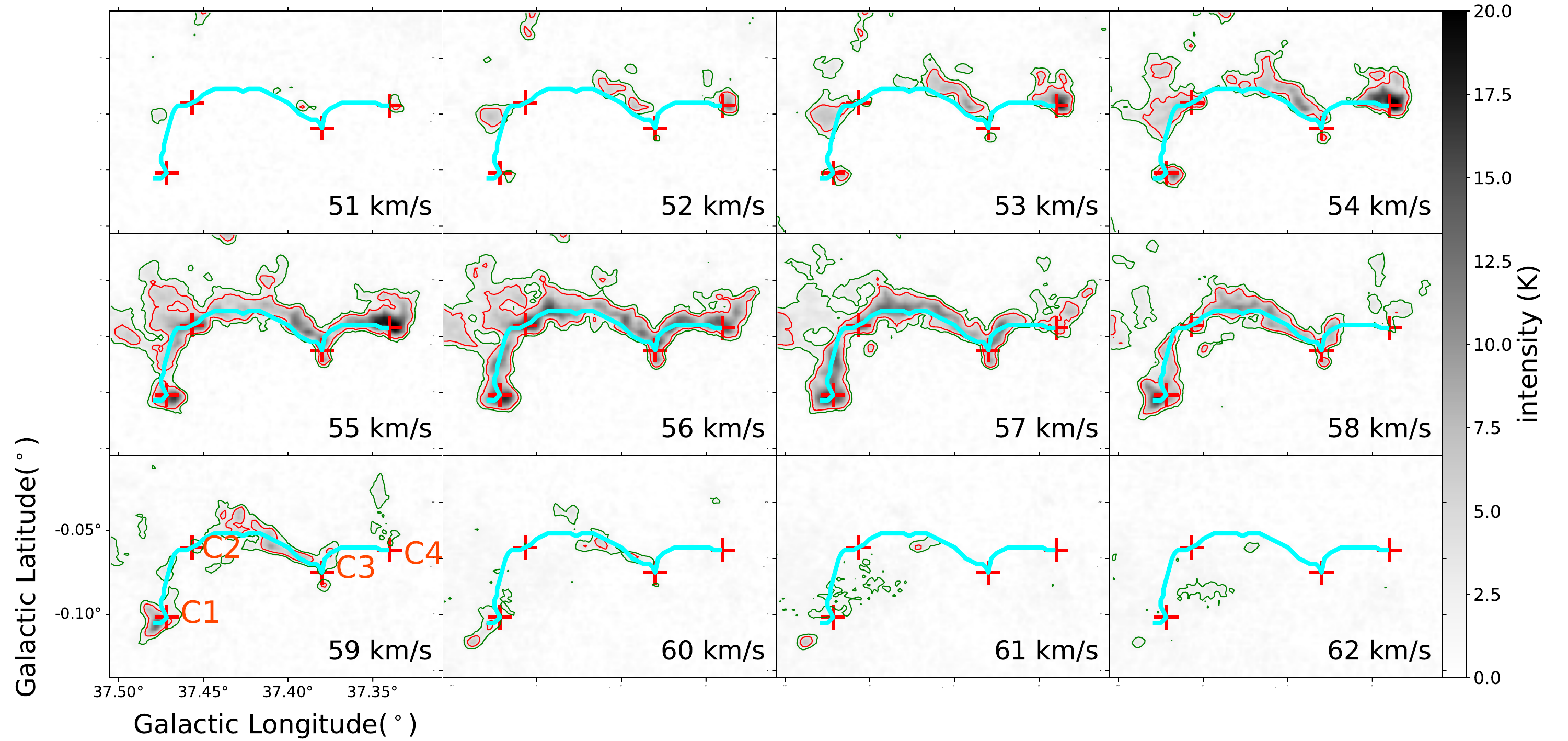}
\caption{Channel maps of $^{12}$CO and $^{13}$CO\,(3-2) emissions are presented for filament G37. The intensity of the 
$^{12}$CO and $^{13}$CO\,(3-2) lines within the velocity range of 51 to 62\,km\,s$^{-1}$ is depicted by green and red contours, respectively, 
with a step of 1\,km\,s$^{-1}$. These contours represent regions where the signal-to-noise ratios of $^{12}$CO and $^{13}$CO 
exceed 6$\sigma$ of the spectral lines, corresponding to approximately 1.9 and 2.1\,K\,km\,s$^{-1}$, respectively. The intensity distribution
of $^{12}$CO emission for each velocity channel is displayed as the gray background of the maps. The skeleton of filament G37 is 
represented by a cyan line, while four special dense clumps (C1--C4) are denoted by red crosses, which are consistent with those shown 
in Fig.\,\ref{fig1}.}
\label{fig2}
\end{figure*}

\section{Archival Data}\label{sect2}
\subsection{Spectroscopic Data}
Carbon monoxide (CO) serves as an excellent tracer for studying the kinematic properties of molecular clouds. The spectroscopic data utilized 
in this study is derived from the CO High-Resolution Survey (COHRS), specifically the $^{12}$CO\,($J$\,=\,3-2) transition, which was observed 
by the James Clerk Maxwell Telescope (JCMT) \citep{Dempsey2013}. The survey covers a region with |$b$|\,$\leq$\,0.5$^\circ$ and 
10.25$^\circ$\,$\lesssim$\,$l$\,$\lesssim$\,56$^\circ$. The beam size of the observations is $\sim$15$''$ with a velocity resolution of
1\,km\,s$^{-1}$ and a pixel size of about 6$''$. The average root mean square (RMS) noise level is $\sim$1\,K. Additionally, this study
incorporates the $^{13}$CO and C$^{18}$O\,($J$\,=\,3-2) emissions from the $^{13}$CO/C$^{18}$O\,($J$\,=\,3-2) Heterodyne Inner Milky Way Plane
Survey (CHIMPS), which was observed as part of the Heterodyne Array Receiver Program of JCMT \citep{2016MNRAS.456.2885R}. The CHIMPS survey
covers a region with |$b$|\,$\leq$\,0.5$^\circ$ and 28$^\circ$\,$\lesssim$\,$l$\,$\lesssim$\,46$^\circ$. The spatial resolution for the 
$^{13}$CO and C$^{18}$O emissions is $\sim$14$''$ with a velocity resolution of 0.5 km\,s$^{-1}$ and a pixel scale of 7$''$. 
The median RMS noise level is $\sim$0.6\,K.

The ionized gas is effectively traced by the H$\alpha$ radio recombination line (RRL). For this investigation, we utilized the 
H$\alpha$\,RRL data obtained from the Green Bank Telescope (GBT) Diffuse Ionized Gas Survey (GDIGS; \citealt{2021ApJS..254...28A}). 
The H$\alpha$\,RRL emission was quantified through the measurement of the 4-8\,GHz radio recombination line. The beam size employed 
was $\sim$2.8$'$, with a velocity resolution of 0.5\,km\,s$^{-1}$.

\subsection{Continuum Data}
In this work, continuum data is used at different wavelengths.
The 850\,$\mu$m continuum emission is observed by JCMT telescope (JCMT Plane Survey, \citealt{2017MNRAS.469.2163E}), of which resolution is around 14$''$.
The continuum at mid-infrared is observed by {\it Spitzer} at 3.6, 8, and 24\,$\mu$m ({\it Spitzer}, GLIMPSE \citealt{2003PASP..115..953B}; MIPSGAL \citealt{2009PASP..121...76C}) {\mk with the resolutions as 1.2$\arcsec$, 2$\arcsec$, and 6$\arcsec$, respectively.}
The continuum at 1420\,MHz is observed by Very Large Array (VLA) from VGPS survey \citep{2006AJ....132.1158S}, which resolution is around 44$''$.

\subsection{Column Density and Dust Temperature }
The H$_2$ column density and dust temperature maps\footnote{\url{https://dataverse.harvard.edu/dataverse/Galactic-Filaments}} of filament G37
come from \cite{2018ApJ...864..153Z}, which is derived by SED fitting procedure \citep{2015MNRAS.450.4043W} on the {\it Herschel} continuum at
wavelengths of 70, 160, 250, 350, and 500\,$\mu$m.

\subsection{Polarization Data}
The 353\,GHz dust polarized emission, obtained from the $Planck$ satellite\footnote{\url{http://www.esa.int/Planck}}
\citep{2020A&A...641A..12P,2020A&A...641A..11P}, offers a valuable means to investigate the large-scale magnetic field structure of molecular 
clouds, as demonstrated in previous studies \citep{2016A&A...586A.138P}. 
Utilizing observations from the High-Frequency Instrument (HFI; \citealt{2020A&A...641A...3P}), 
researchers have generated maps of the Stokes parameters $I$, $Q$, and $U$, along with their corresponding dispersion values 
($\sigma I$, $\sigma Q$, $\sigma U$). These maps possess a resolution of 5$'$ and a pixel size of $\sim$1.7$'$. 
The polarization angle can be derived from the HFI Stokes maps, $\psi_{\rm {Planck}}\,=\,0.5\,\times\,{\rm {arctan}}(U,Q)$,
where $\psi_{\rm Planck}$ varies from -90$^\circ$ to 90$^\circ$ with the HEALPix convention. 
To align with the IAU convention, the $Planck$ measurement needs to be converted using the equation $\psi$=\,0.5\,$\times$\,arctan(-$U$,$Q$). 
To ascertain the orientation of the magnetic field (referred to as the $B$-field), one can obtain $\psi_{\rm B}$ by adding 90$^\circ$ to the 
polarization angle. This can be achieved using the equation $\psi_{\rm B}$\,=\,$\psi_{\rm Planck}$\,+\,90$^\circ$.

\section{Result}
\label{sect3}
\subsection{Overview}
\label{Overview}
Previous studies conducted by \cite{2016A&A...591A...5L}, \cite{ 2016ApJS..226....9W}, \cite{2016MNRAS.456.2885R}, and \cite{2018ApJ...864..153Z} 
have identified and characterized filament G37, revealing a velocity range of 51 
to 63\,km\,s$^{-1}$. 
Utilizing the BeSSeL Survey calculator \citep{2019ApJ...885..131R}, it has been determined around a 50 percent probability that the filament G37 is situated in the Sagittarius far arm, approximately 9.3$\pm$0.5\,kpc away. The length of filament G37 measures $\sim$41$\pm$2\,pc.

As depicted in Fig.\,\ref{fig1}, the 850\,$\mu$m emission of filament G37 reveals a distinct shape reminiscent of a caterpillar, 
featuring two semicircular structures within its main body. \cite{2018ApJ...864..153Z} have identified four dense clumps along 
the skeleton of filament G37. These clumps, namely C1, C2, C3, and C4, correspond to the eastern endpoint of filament G37, 
the peak position of the larger semicircular structure, the junction point between the two semicircular
structures, and the western endpoint of filament G37, respectively. These clumps exhibit higher brightness in the 850\,$\mu$m 
emission compared to other regions of the filament. Additionally, Fig.\,\ref{fig1} displays bright emissions at 3.6, 8, and 24\,$\mu$m 
in these clumps, indicative of ongoing star formation activities resulting from the heating of dust by newly formed stars. 
Therefore, these clumps are likely sites of active star formation.

Surrounding the filament G37, emissions at 8\,$\mu$m occur within two semicircle structures, as depicted in Fig.\,\ref{fig1}. Additionally, 
the 1420\,MHz and H$\alpha$\,RRL emissions were observed in the same region. These emissions suggest the existence of H\,{\scriptsize II} regions 
in the vicinity of the filament G37.

Based on the Spicy Survey \citep{2021ApJS..254...33K} and Gaia EDR3 \citep{2021AJ....161..147B} catalog, we found multiple Young Stellar Object (YSO) candidates associated with filament G37 and H\,{\scriptsize II} regions. 
We selected the YSOs from the Spicy Survey catalog that are located at distances greater than 5\,kpc, utilizing the Gaia database. This selection criterion is based on the filament's distance, which is approximately 9.3\,kpc.
Notably, some YSO candidates are found within the filament body itself and near the four 
dense clumps, as illustrated in Fig.\,\ref{fig1}. These findings suggest the potential for star formation within filament G37.

Fig.\,\ref{fig1} illustrates the spatial distribution of H$_2$ column density ($N$(H$_2$)) and dust temperature ($T_{\rm dust}$) within 
filament G37, as determined through SED fitting of far-infrared continuum observations obtained by the {\it Herschel} telescope at wavelengths 
of 70, 160, 250, 350, and 500\,$\mu$m \citep{2018ApJ...864..153Z}. The column density of filament G37 ranges from 1.0\,$\times$\,10$^{22}$ 
to 2.3\,$\times$\,10$^{22}$\,cm$^{-2}$, with a mean value of $\sim$1.3\,$\times$\,10$^{22}$\,cm$^{-2}$ (see also Table\,\ref{tab1}). 
Notably, the column density 
structure aligns well with the distribution of 850\,$\mu$m emissions (see Fig.\,\ref{fig1}). The four dense clumps (C1, C2, C3, and C4)
exhibit high column density values of $\sim$1.3--1.8\,$\times$\,10$^{22}$\,cm$^{-2}$.

Regarding the dust temperature, filament G37 displays a range of 15 to 24\,K. The dense clumps C1, C2, C3, and C4 exhibit slightly lower dust 
temperatures, averaging around 20\,K. Conversely, the regions within the two semicircle structures generally exhibit higher dust temperatures, 
ranging from 24 to 35\,K. This temperature contrast between the interior and exterior of filament G37 is significant.

\subsection{Distributions of CO and H$\alpha$\,RRL}
In this study, we employ the $^{12}$CO, $^{13}$CO, C$^{18}$O, and H$\alpha$\,RRL emissions to investigate the dynamical properties 
of filament G37. Three mean spectral lines of $^{12}$CO, $^{13}$CO, and H$\alpha$\,RRL observed in filament G37 are illustrated in Fig.\,\ref{fig7}.  
The identification of filament G37 and determination of its velocity range as 51--63\,km\,s$^{-1}$ were 
accomplished by \cite{2018ApJ...864..153Z}. To visualize the gas morphology within the filament, intensity maps of the $^{12}$CO, 
$^{13}$CO, and H$\alpha$\,RRL emissions at velocity of 51--63\,km\,s$^{-1}$ are presented in Fig.\,\ref{fig1}.

In the central region of the filament G37 body, a notable agreement is observed between the dust emission at 850\,$\mu$m and 
the emissions of $^{12}$CO and $^{13}$CO within the primary section of the filament, as depicted in Fig.\,\ref{fig1}. 
However, a discrepancy arises between the 850\,$\mu$m and $^{12}$CO, $^{13}$CO emissions towards the southern portion of 
a small semicircle structure associated with filament G37. Specifically, a clump ($V_{\rm LSR}$(CO)\,=\,$\sim$42\,km\,s$^{-1}$) is detected 
in the 850\,$\mu$m continuum, 
but it does not correspond to the intensity map of the $^{12}$CO and $^{13}$CO emissions at velocity range of 51--63\,km\,s$^{-1}$. 
This discrepancy suggests that the clump and filament G37 may not occupy the same position along the line of sight (LOS) direction.

The distribution of diffuse gas, as traced by the $^{12}$CO\,(3-2), is observed in the northeastern and northwestern regions of 
filament G37 (see Figs.\,\ref{fig1} and \ref{fig2}). Notably, the highest integrated intensity values from both $^{12}$CO and $^{13}$CO 
emissions are found at the extremities of filament G37. Furthermore, within this region, two dense clumps are identified. 
The first clump, G37.341-00.062, is associated with young star objects identified through the ATLASGAL survey \citep{2014MNRAS.443.1555U}, 
as depicted in Fig.\,\ref{fig1}. The second clump corresponds to the H\,{\scriptsize II} region G37.469-0.104, as cataloged in the 
WISE catalog \citep{2014ApJS..212....1A}, also shown in Fig.\,\ref{fig1}.

\begin{figure}
\centering
\includegraphics[width = 9cm]{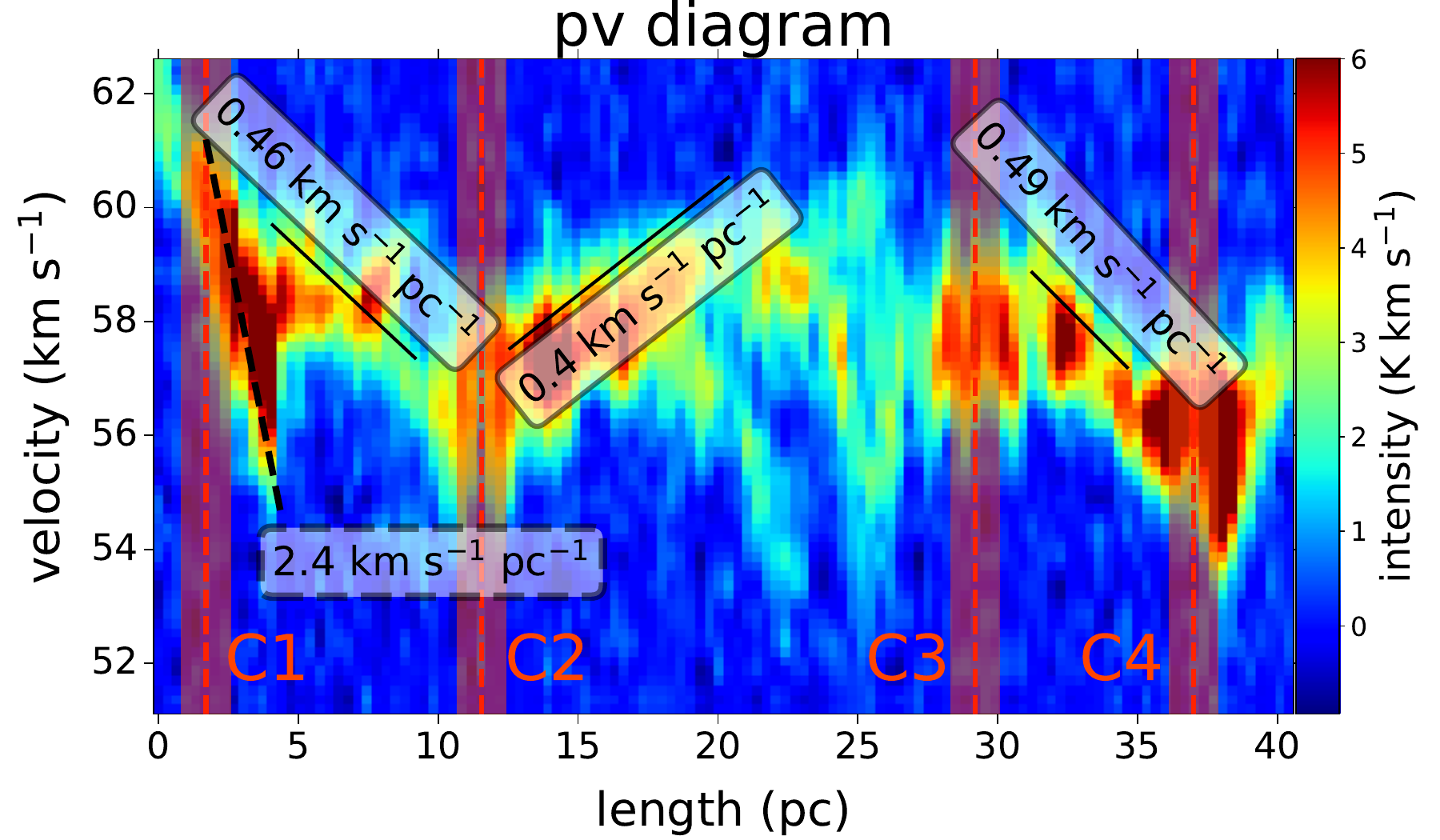}
\caption{The position-velocity (P-V) diagram is presented to illustrate the distribution of filament G37, which was obtained from $^{13}$CO 
emission. The distribution is shown along the filament skeleton, as depicted in Fig.\,\ref{fig2}, with the dense clumps C1, C2, C3, and C4 
(as shown in Fig.\,\ref{fig1}) represented by the red columns. The x-axis denotes the filament length in pc units, while the y-axis shows 
the velocity $V_{\rm LSR}$. The diagram shows red vertical dashed lines for the locations of four dense clumps (C1--C4) and black lines for the velocity gradients. The extra velocity gradient of clump C1 is shown as a dashed line.}
\label{fig3}
\end{figure}

The H$\alpha$\,RRL emissions exhibit similarities to the 1420\,MHz emission (see Fig.\,\ref{fig1}). The occurrence of multiple H$\alpha$\,RRL 
emissions surrounding the two semicircles of the filament G37 indicates the presence of H\,{\scriptsize II} regions in close 
proximity to the filament. In the H\,{\scriptsize II} region denoted as G37.469-0.104 \citep{2014ApJS..212....1A}, weak emissions of 
H$\alpha$\,RRL and 1420\,MHz continuum are observed. This implies that G37.469-0.104 potentially signifies a compact H\,{\scriptsize II} region.

The estimation of the excitation temperature for $^{12}$CO\,(3-2) can be achieved through the application of local thermodynamic equilibrium 
(LTE) as described by \cite{2014A&A...572A..56P},
\begin{equation}
    \begin{aligned}
        T_{\rm ex}(^{12}{\rm CO}\,3-2)\,=\,\frac{16.59}{{\rm ln}[1+16.59/\,(T_{\rm mb}+0.036)]} ~~{\rm K}, 
    \end{aligned}
\end{equation}
where $\it{T}_{\rm mb}$ is the peak temperature of $^{12}$CO (3-2) emission. 
Fig.\,\ref{fig1} illustrates a complex distribution of excitation temperatures within filament G37. It is observed that the excitation 
temperature of the predominant gas within the filament ranges from 25 to 41\,K.
The excitation temperature is not uniformly distributed along the filament body. Notably, clumps C2 and C4 exhibit elevated temperatures exceeding 30\,K (see Fig.\,\ref{fig1}). Additionally, outside the semicircular structure, the area adjacent to clump C2 also displays a high excitation temperature. In contrast, the excitation temperatures of other regions along the filament are comparable to the dust temperature.

\subsection{Velocity Field}\label{sec3.4}
Fig.\,\ref{fig2} illustrates the channel maps of $^{12}$CO and $^{13}$CO\,(3-2) of filament G37, spanning from $V_{\rm LSR}$\,=\,51 to 62\,km\,s$^{-1}$. 
The gas structures observed in the $^{12}$CO and $^{13}$CO\,(3-2) emissions exhibit similar characteristics in each velocity channel. 
The molecular tracer $^{12}$CO reveals more detailed structures within the diffuse molecular gas compared to the 
$^{13}$CO emission. Initially, at velocity from 51 to 54\,km\,s$^{-1}$, the dense clump C4 (G37.341-00.062) emerges, followed by the 
eastern top of the large semicircle (Clump C2) and the connecting part between two semicircles (Clump C3). The molecular gas exhibits 
an east-to-west flow along the filament structure within the small semicircle, towards the dense clump C4. Between $V_{\rm LSR}$\,=\,53 
and 55\,km\,s$^{-1}$, the gas within the two semicircle structures of the filaments gradually converges towards the connecting part 
(Clump C3). The gas within the large semicircle flows from the top part (Clump C2) towards the two sides (Clumps C1 and C3). 
Subsequently, from $V_{\rm LSR}$\,=\,54 to 59\,km\,s$^{-1}$, the gas flows from west to east, following the filament structure. 
From $V_{\rm LSR}$\,=\,59 to 62\,km\,s$^{-1}$, only the Clump C1 remains, gradually diminishing. 

The position-velocity (P-V) diagram depicted in Fig.\,\ref{fig3} corresponds to the filament skeleton, as illustrated in Fig.\,\ref{fig1}.
Material flows predominantly occur within the filament, as evidenced by the channel maps and P-V diagram presented in Figs.\,\ref{fig2} 
and \ref{fig3}, respectively. Notably, significant velocity gradients are observed between dense clumps C1 and C2, as well as between C3 and C4, 
at approximately 0.46 and 0.49\,km\,s$^{-1}$\,pc$^{-1}$, respectively. Between dense clumps C2 and C3, the left portion of the filament 
exhibits a distinct velocity gradient of $\sim$0.4\,km\,s$^{-1}$\,pc$^{-1}$, while the velocity distribution in the other half remains unclear.
In the vicinity of dense clump C1 within a radius of less than 5\,pc, a significant velocity gradient is detected, as illustrated in Fig.\,\ref{fig3}, exhibiting a wide range of velocities ($\sim$\,2.4\,km\,pc$^{-1}$, see Fig.\,\ref{fig2}). A similar result has been found in 
the OMC-1 region, where a velocity gradient of 5--7\,km\,pc$^{-1}$ was measured \citep{2017A&A...602L...2H}. This phenomenon may indicate the 
presence of accelerated motions towards the massive dense clump C1.

Fig.\,\ref{fig4} illustrates the distribution of the central velocity ($V_{\rm LSR}$) along the line-of-sight and the velocity dispersion
($\sigma$V) for both $^{12}$CO and $^{13}$CO\,(3-2) emissions. The coverage region of the central velocity and 
velocity dispersion maps derived from $^{13}$CO\,(3-2) is limited to areas where the signal-to-noise ratios (SNRs) exceed 2$\sigma$ (5\,K\,km\,s$^{-1}$). 
In comparison to the $^{13}$CO, the SNRs of the $^{12}$CO emission are relatively high. 
In Fig.\,\ref{fig4}, the coverage region of $^{12}$CO exceeds 5$\sigma$ (5.5\,K\,km\,s$^{-1}$).
As depicted in Fig.\,\ref{fig4}, the velocity distribution within the western region of the filament is relatively low, 
while it is higher in the eastern region. Furthermore, the velocity dispersion of $^{13}$CO is lower compared to that of $^{12}$CO. 
The diffuse gas surrounding the filament displays a higher velocity dispersion, as measured by the $^{12}$CO spectral line.

\begin{figure}
\centering
\includegraphics[width = 9.5cm]{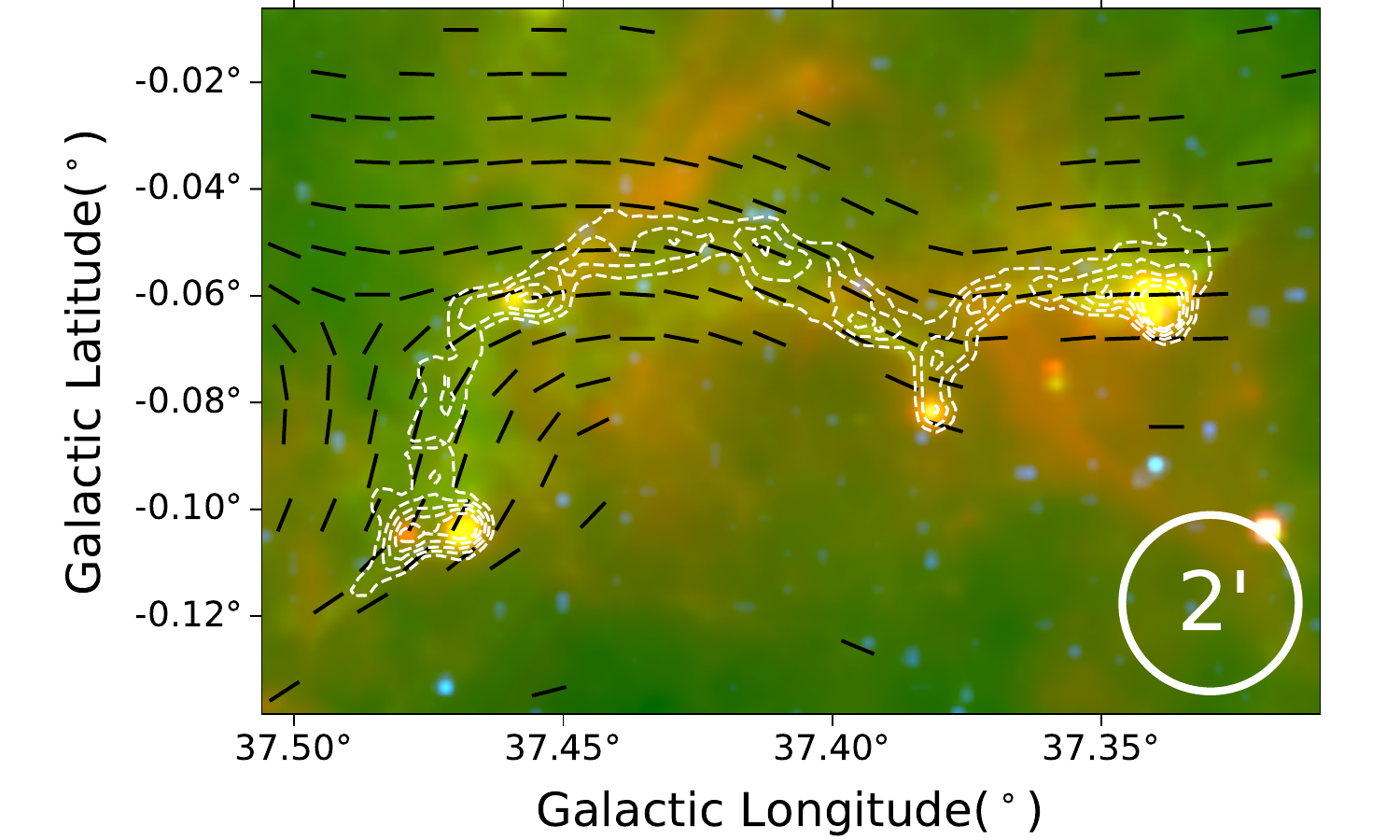}
\caption{The magnetic field structure of filament G37 is revealed with the VGT measurements of $^{12}$CO. The orientation of the magnetic field 
is obtained and represented as black vectors. The intensity map of filament G37, derived from $^{13}$CO emission, is depicted as white 
contours (same as Fig.\,\ref{fig1}). The background of the image corresponds to a mid-infrared three-color map, consistent with 
Fig.\,\ref{fig1}. The spatial resolution of the magnetic field measurement using VGT is $\sim$2$'$, as indicated by a white circle 
in the bottom right corner.}
\label{fig6}
\end{figure}

\begin{table}[]
    \centering
    \caption{Physical parameters of filament G37 }
    \begin{tabular}{c c c}
    \hline
	\hline
	Property & Filament G37 & Units \\
	\hline
    Effective length & 41 & pc \\
    Effective radius & 3.0 & pc \\
    1D Velocity dispersion & 1.3 & km\,s$^{-1}$ \\
    H$_2$ Column density & 1.3 & $\times 10^{22}$\rm cm$^{-2}$ \\
    H$_2$ Volume density & 737 & cm$^{-3}$ \\
    Volume mass density & 3.5 & $\times10^{-21}$ g cm$^{-3}$ \\
    Dust temperature & 24 & K \\
    Sound speed & 289 & m\,s$^{-1}$ \\
    Sonic Mach number & 7.6 & \\
    Alfv\'{e}n number & 2.0 & \\
    Magnetic field strength & 23.3 & $\mu$G \\
    \hline
    \hline
    \end{tabular}
   \tablefoot{Effective length and radius of filaments were measured with Python Package: RadFil \citep{2018ApJ...864..152Z}.}
    \label{tab1}
\end{table}

\subsection{Magnetic Field}
\label{sect3.6}
As mentioned in Sect.\,\ref{Overview}, filament G37 is situated in the Sagittarius far arms, $\sim$9.2\,kpc away, within the Galactic plane. 
The measurement of the magnetic field of filament G37 poses a challenge due to foreground and/or background effects, 
as it traverses multiple spiral arms along the line of sight.
A new technique, Velocity Gradient Technique (VGT), can use the magnetohydrodynamic (MHD) turbulent anisotropy from position-position-velocity cube to infer magnetic field \citep{2019NatAs...3..776H,2022ApJ...934...45Z,2024ApJ...961..124Z,2024ApJ...967...18Z}.
Therefore, we employ the VGT (\citealt{2017ApJ...835...41G,2018ApJ...853...96L,2018MNRAS.480.1333H,2024ApJ...961..124Z,2024ApJ...967...18Z}) to measure the magnetic field, 
which mitigates the influence of foreground and/or background effects. The description and accuracy of the VGT are elaborated in
Appendix\,\ref{VGTm}.
The application of VGT analysis to all velocity components can validate the accuracy of VGT in relation to the magnetic field within this region (see Fig.\,\ref{fig5}). By utilizing the velocity component of filament G37, we can delineate the magnetic field structure of G37 clearly, free from the influences of foreground and background effects (see Appendix\,\ref{VGTm}).

The velocity dispersion at each pixel exceeds 1\,km\,s$^{-1}$ (velocity channel width), indicating that the velocity channels derived from
the position-position-velocity (PPV) cube of $^{12}$CO and $^{13}$CO\,(3--2) emissions are narrow \citep{2018ApJ...853...96L}. This narrow velocity 
channel can effectively trace the turbulent velocity field \citep{2001ApJ...555..130L}. 
The special lines of $^{12}$CO exhibit a high signal-to-noise ratio, enabling the differentiation of various velocity components within the G37 filament region (see Fig.\,\ref{fig5}). In contrast, the signal-to-noise ratio of $^{13}$CO is insufficient to validate the accuracy of the VGT for tracing the magnetic field (see Fig.\,\ref{fig7}).
To measure the magnetic field of filament G37, 
we applied the VGT to the $^{12}$CO spectral line at a velocity range of 51--64\,km\,s$^{-1}$ (see Sect.\,\ref{Overview}), 
corresponding to velocity component 4 of filament G37, as shown in Fig.\,\ref{fig5}. The magnetic field structure of filament G37 
is presented in Fig.\,\ref{fig6}. The magnetic field resolution is $\sim$2$'$, and the sub-block size was set to 20$\times$20 pixels. 
In all cases, the magnetic field structure was found to be nearly parallel to the filament G37 (see Fig.\,\ref{fig6}). 
Within the large semicircle of filament G37, the magnetic field orientations were distributed along the filament and aligned 
parallel to its elongation direction. Conversely, within the small semicircle, the magnetic field orientations were parallel 
to the Galactic Plane and aligned in the east-west direction.

\subsection{Identification of Clumps}
\label{clumps}
Multi-velocity components exist around the filament in the line-of-sight orientation (see Figs.\,\ref{fig7}, \ref{fig5}, and Sect.\,\ref{sect3.6}).
The velocity-integrated intensity maps of these components are presented in Fig.\,\ref{figE}. The H$_2$ column density obtained from the {\it Herschel} \citep{2018ApJ...864..153Z} exhibits a structure similar to that of component 4, which corresponds to a velocity range of 51--63\,km\,s$^{-1}$ and delineates the G37 filament. In contrast, the other components display weak $^{12}$CO emissions and diffuse distributions, partially overlapping with the G37 filament. Consequently, the emissions observed from {\it Herschel} and the 850\,$\mu$m continuum are likely primarily associated with the G37 filament.
To identify the location of clumps in filament G37, we employed $^{13}$CO\,(3-2) and 850 $\mu$m emissions. 
The clumps found in the 850 $\mu$m emission that corresponded with the velocity range 51-62\,km\,s$^{-1}$ of $^{13}$CO 
emission were considered to be situated on filament G37. We identified 17 dense clumps that are located within the filament G37  (details see Appendix\,\ref{ApD}). 
The positions of clumps are shown in Fig.\,\ref{B1} and Table\,\ref{tab2}. We assume that each clump is a uniform sphere 
\citep{2000MNRAS.311...85F}. The effective radius $R$ of clumps can be estimated by, $R=\sqrt{A/\pi}$, where the $A$ is 
the coverage area of clumps. The clump size is the effective diameter ($D=2R$). 
Mass of clumps $M$ is calculated by, $M=N(\rm {H_2})\,\mu m_{\rm H} A$,
where $N$(H$_2$) is H$_2$ column density, $\mu$ is the mean molecular weight for clouds as 2.37, $m_{\rm H}$ is the mass of H atom. 
The details of the parameters of dense clumps are shown in Table\,\ref{tab2}. 

\begin{figure}
\centering
\includegraphics[width=9cm]{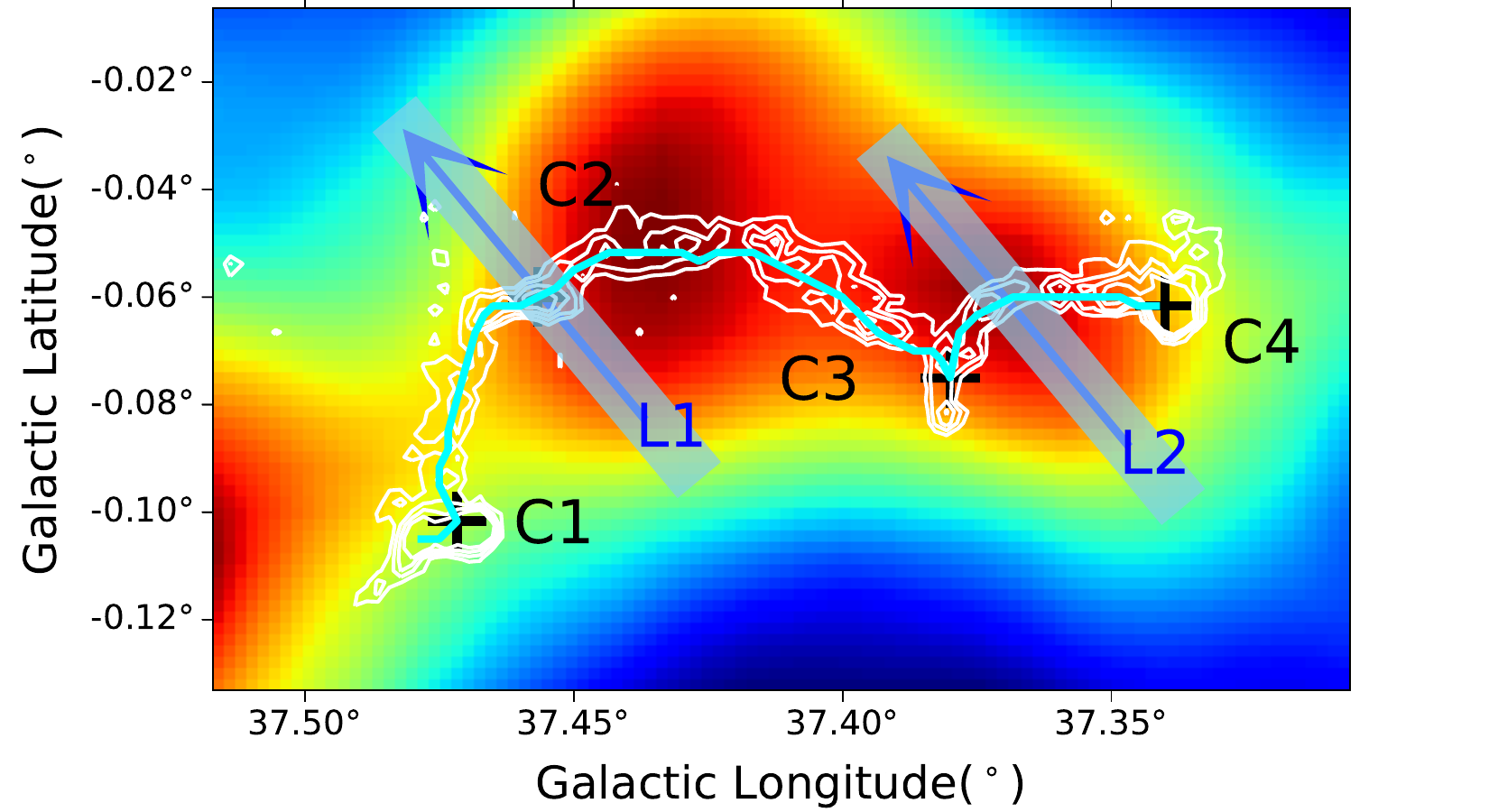}
\includegraphics[width=9cm]{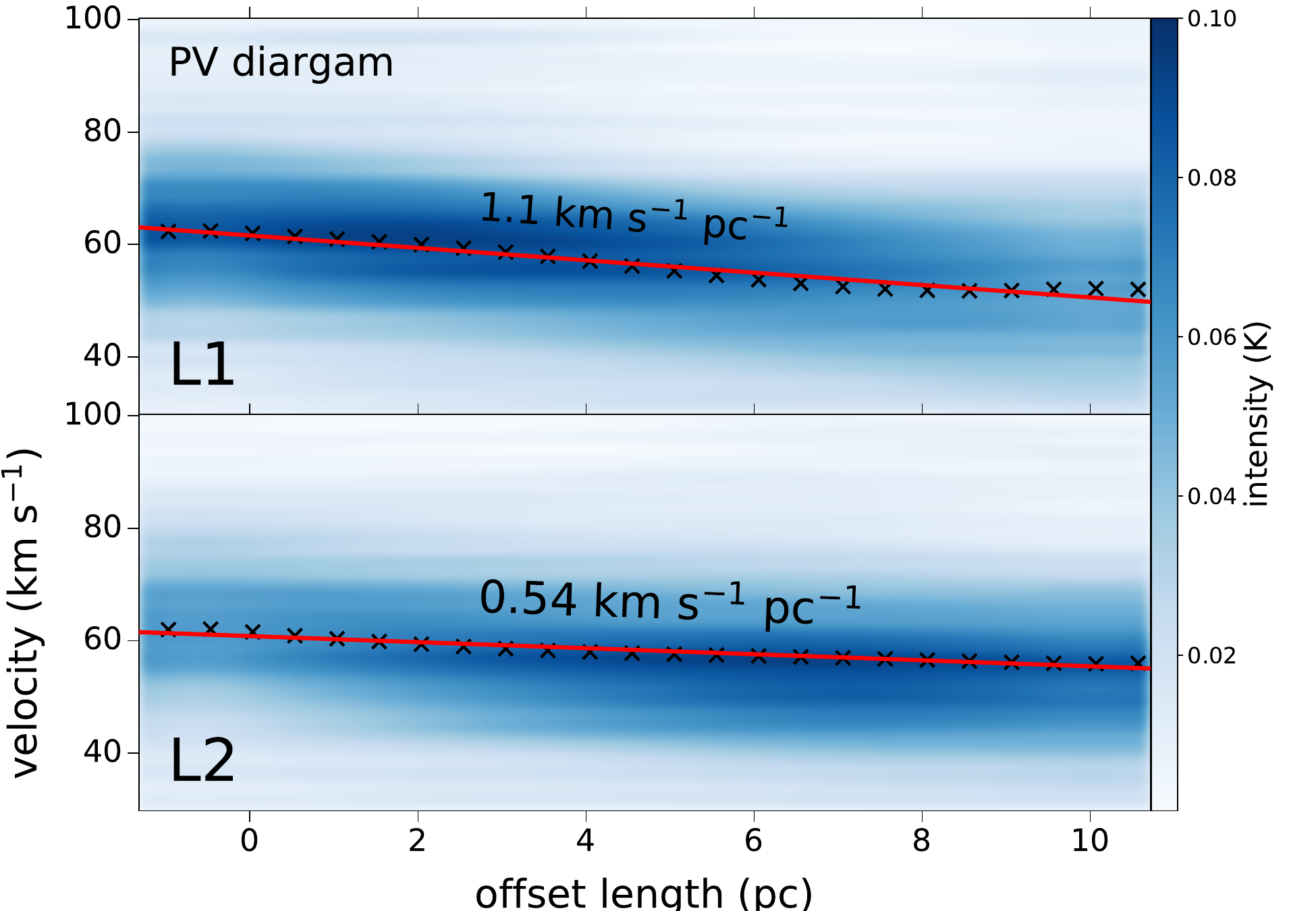}
\caption{The P-V diagram obtained from H$\alpha$\,RRL observations is presented. In the top panel, the intensity map of H$\alpha$\,RRL emission
is displayed in the background, with the filament shape depicted by white contours (same as Fig.\ref{fig1}). The filament skeleton is
represented by a cyan line. Four key points of filament G37, as shown in Fig.\ref{fig1}, are marked by black crosses. The P-V diagrams 
are derived from two blue lines, namely L1 (middle panel) and L2 (bottom panel), which pass through the top points of two semicircles. 
The velocity gradient of H$\alpha$\,RRL emission is indicated by red lines in the bottom panels, obtained by fitting the weighted velocity
positions (refer to the black cross in the middle and bottom panels).}
\label{fig8}
\end{figure}

The timescale that reflects the evolutionary features of a filament is the time it takes to grow to reach its critical line mass ($M_{\rm cri}$). 
This timescale is known as the critical timescale ($\tau_{\rm cri}$) and represents the lower age limit of the filament. 
The function of the fragmentation length scale can be used to calculate the critical timescale $\tau_{\rm cri}$ \citep{2018A&A...613A..11W},
$\rm \tau_{age}\leq\tau_{cri}\approx\frac{\lambda_{core}}{2c_s}$,  where the $\lambda_{\rm core}$ represents the separation between the clumps/cores in a filament. 
The length between nearby clumps in the filament G37, $\lambda_{\rm core}$, is roughly estimated at around 3\,pc due to the low spatial resolution. 
{\mk The critical timescale ($\tau_{\rm cri}$) for filament G37 is estimated to be approximately 4.9\,Myr. However, this estimation is subject to considerable uncertainties arising from several factors, including limited spatial resolution, distance measurement errors, and projection effects. Notably, the separation between clumps ($\lambda_{\rm core}$) within G37 is highly uncertain. Furthermore, the assumption of uniform temperature variations introduces additional errors into the estimation. As a result, the critical timescale of $\tau_{\rm cri}$\,$\sim$\,4.9\,Myr should be regarded as a lower limit for filament G37, accompanied by a degree of uncertainty.}



\begin{figure*}[h]
\centering
\includegraphics[height = 7cm]{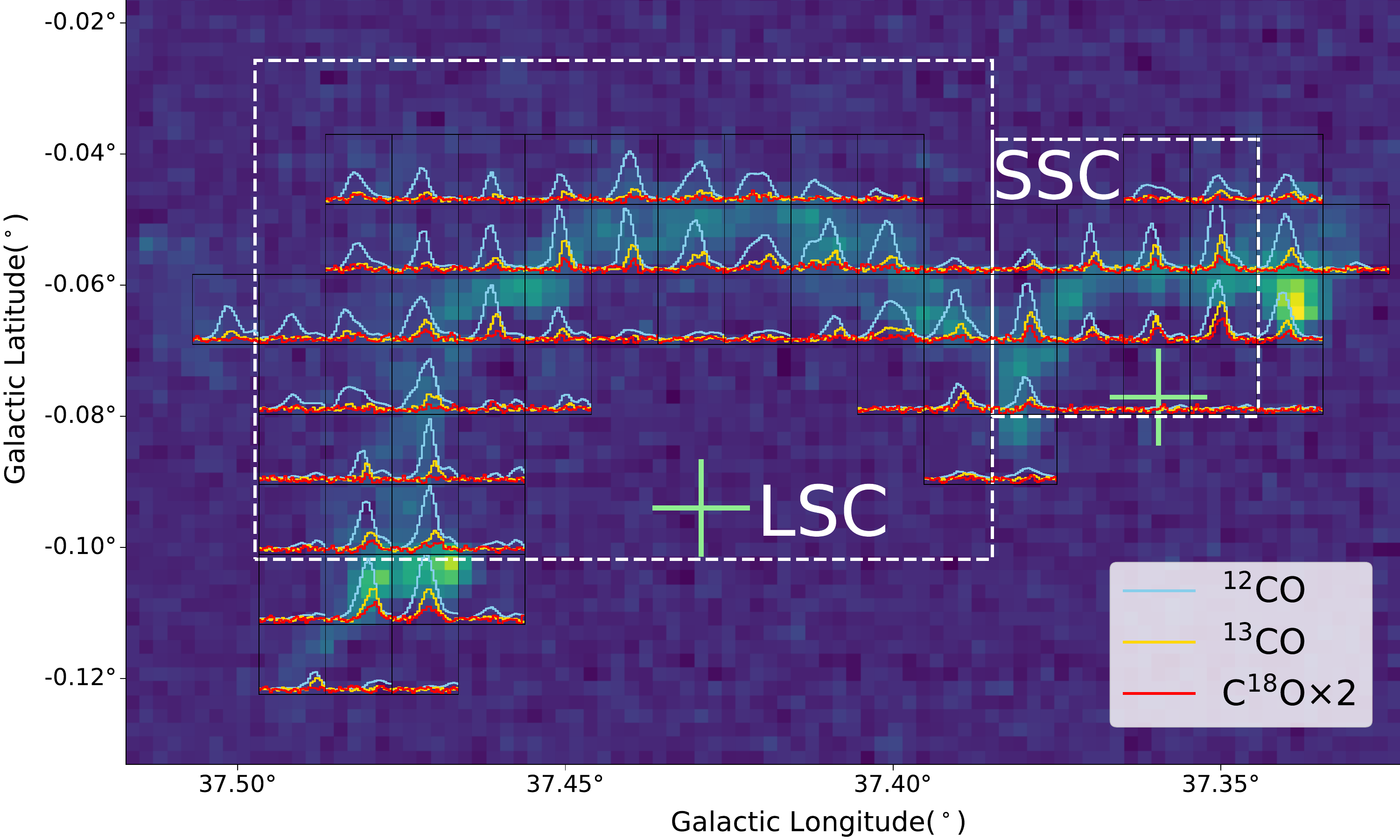}
\includegraphics[height = 7cm]{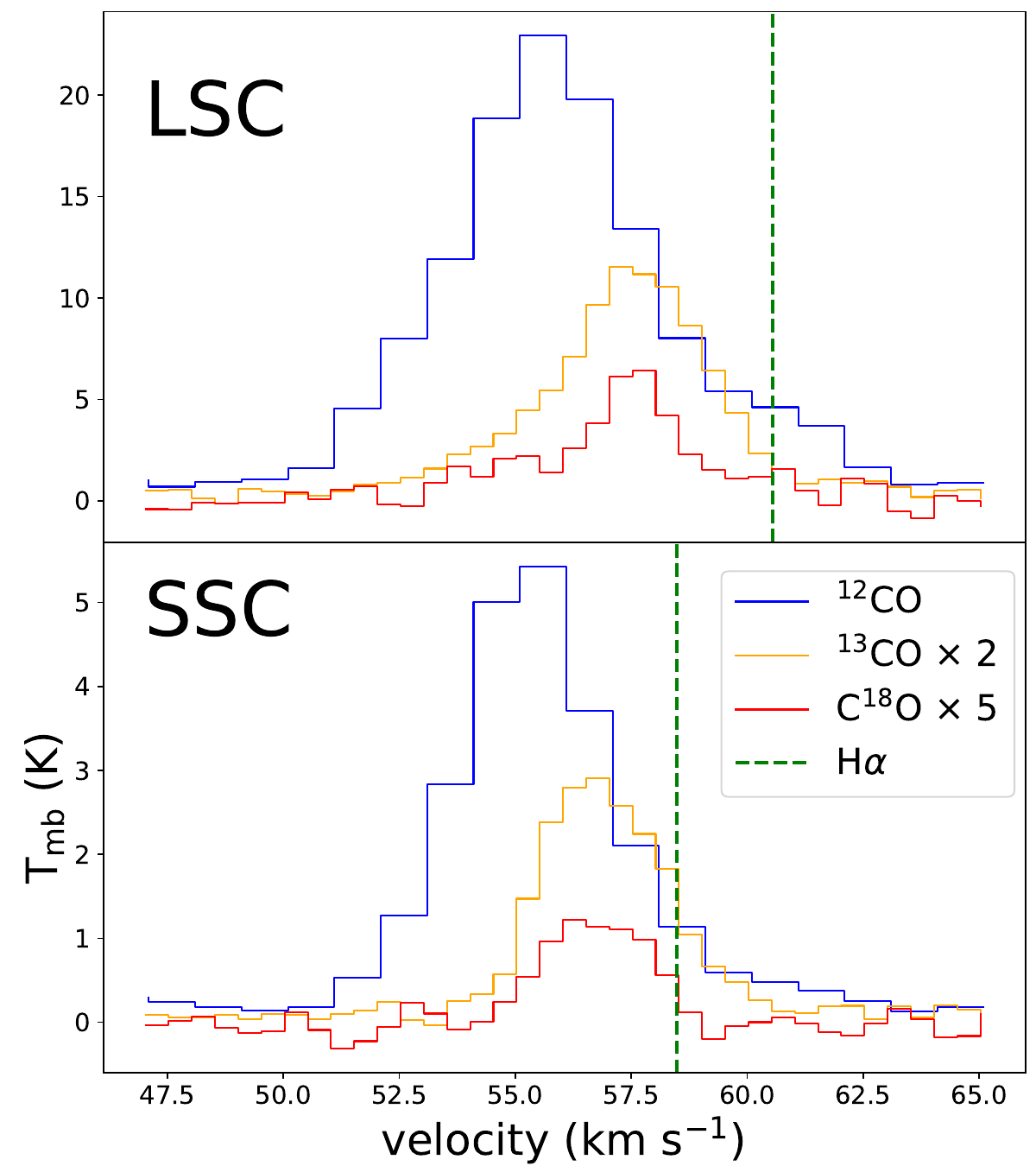}
\caption{The spectral line shapes of four tracers are depicted in filament G37. 
The spectra contours of $^{12}$CO, $^{13}$CO, and C$^{18}$O\,(3-2) are represented by the blue, yellow, and red lines, respectively, within each block (30$''\times$30$''$). 
The velocity range for these lines is from 47 to 65\,km\,s$^{-1}$. 
The background displays the intensity map of $^{13}$CO (3-2) at velocity range of[47,65] km\,s$^{-1}$.
The right panels display the spectral lines of the large semicircle (LSC) and small semicircle (SSC), which cover the region shown in the left panel. 
The central velocities of the H$\alpha$\,RRL lines, indicated by the green cross in the left panel, 
are marked by dotted vertical lines. Each spectral line is distinguished by its respective color.}
\label{fig9}
\end{figure*}

\section{Discussion}
\label{sect4}
\subsection{Material-Flows and End Collapse}
\label{Material-Flows}
Filament G37 is a unique sample that distinguishes itself from other giant filaments located in the Galactic plane. 
What sets it apart is not only its curved configuration but also its presence within multi-directional material flows.
While the curvature of filaments in the Galactic plane is not uncommon, such as seen in G51, G11, G47, IC\,446/IC\,447, 
NGC\,6334 \citep{2013A&A...559A..34L,2013A&A...554L...2Z,2015MNRAS.450.4043W,2020ApJ...899..167B}, 
filament G37 diverges in its characteristics. Unlike the velocity structure of these giant filaments \citep{2024ApJ...961..124Z}, 
multiple velocity gradients (around 0.5 km\,s$^{-1}$\,pc$^{-1}$) are found in filament G37 (see Fig.\,\ref{fig3}), in which the 
material flows from the top of a large semicircle to the end of this structure and another flow from the connecting point of 
two semicircle structures to the end of the filament body. The material flows along the two semicircle structures of the filament 
G37 and exhibits velocity gradients in multiple directions (details see Sect.\,\ref{sec3.4}). The obvious feature of multi-directional 
material flows occur at the top point of the large semicircular structure where the flow direction of material changes (see Fig.\,\ref{fig3}). 
This phenomenon was also observed in filament S242 \citep{2019ApJ...877....1D,2020A&A...637A..67Y}.

The gravitational instability of filaments can be determined by the critical line mass $M_{\rm cri}$, which is reached when the line mass 
is above the critical value. The critical line mass can be estimated using the kinetic temperature ($T_{\rm k}$) \citep{1964ApJ...140.1056O}:
\begin{equation}
    \begin{aligned}
        M_{\rm cri} = \frac{2 c_{\rm s}^2}{G} \approx \frac{16{\rm M_\sun}}{\rm pc} \left( \frac{T_{\rm k}}{10\,{\rm K}} \right),
    \end{aligned}
\end{equation}
where $c_{\rm s}$ is sound speed. The dust temperature $T_{\rm d}$ is adopted as the kinetic temperature $T_{\rm k}$ of gas in this work.
The critical mass $M_{\rm cri}$ of filament G37 is estimated to be between 30-43\,M$_\sun$\,pc$^{-1}$.
The total mass of the filament G37 is measured to be around 70-135\,M$_\sun$\,pc$^{-1}$, which is much higher than the critical mass.
This suggests that the filament G37 is undergoing gravitational instability. The filament body may break apart, forming clumps.
The initial aspect ratio $A$ of a filament is defined as $A=L/2R$, where $L$ is the filament length and $R$ is the effective radius 
of the filament. If $A$ is less than 5, homologous collapse is observed, whereas if $A$ is greater than 5, terminal-dominated 
collapse is predominant \citep{2011ApJ...740...88P,2012ApJ...756..145P}. The initial aspect ratio $A$ of filament G37 is $\sim$6.7.
End collapse is most likely to occur in filament G37, where H\,{\scriptsize II} regions G37.469-0.104 and dense clump G37.341-0.062 
have been formed at the end of the filament body. The filament G37 provides a potential candidate for end-dominated collapse.


\subsection{Expanding H\,{\scriptsize II} Region}
\label{Expanding-HII}
The morphology of filament G37 exhibits a distinctive feature, consisting of two semicircular structures resembling a "Caterpillar", 
which could be caused by compression from surrounding H\,{\scriptsize II} regions. The extended 1420\,MHz continuum emission 
surrounding filament G37 is the candidate of H\,{\scriptsize II} region affecting filament body (see Fig.\,\ref{fig1}).
H$\alpha$\,RRL has a similar peak velocity to that of filament G37 system velocity traced by $^{12}$CO, $^{13}$CO spectral 
lines (see Fig.\,\ref{fig7}). The H\,{\scriptsize II} regions traced by H$\alpha$\,RRL (see Fig.\,\ref{fig1}) could be close 
to filament G37 in three-dimensional space, which could affect the structure and physical processes of filament G37.
Compared with the inner of the filament G37, the high dust temperature exists regions near the H\,{\scriptsize II} regions 
(see Fig.\,\ref{fig1}), which may present these H\,{\scriptsize II} regions heating the filament body.
These H\,{\scriptsize II} regions could affect the filament body causing the formation of two semicircle structures
of filament G37.

The H\,{\scriptsize II} regions traced by H$\alpha$\,RRL have noticeable velocity gradients when they cross the filament body 
(see Fig.\,\ref{fig8}). These velocity gradients at the top points of large and small semicircle structures are estimated to 
be 1.1 and 0.5\,km\,s$^{-1}$\,pc$^{-1}$ of blue shift, respectively. 
In the two semicircular structures, a notable disparity in blue-shifted velocities is observed between the molecular gas, traced by CO, 
and the ionized gas emanating from the  H\,{\scriptsize II} regions, probed by H$\alpha$\,RRL (see Fig.\,\ref{fig9}). 
The diffuse molecular gas (traced by $^{12}$CO) and the dense gas (traced by $^{13}$CO and C$^{18}$O) exhibit a similar blue-shifted 
velocity difference. These observations serve as direct evidence that the H\,{\scriptsize II} region adjacent to filament G37 is 
expanding towards the blue shift direction, exerting pressure on the primary body of the filament and contributing to its heating 
on a large scale. The multi-directional material flows may be influenced by the expansion process originating from H\,{\scriptsize II} 
regions, with their directional transition point typically situated near the apex of a semicircular structure compressed by the 
expending H\,{\scriptsize II} region.

The timescale of forming semicircle structures on filament G37, $t_{\rm s}$, can be roughly estimated with the physical process
likely caused by the squeezing of the filament body by the H\,{\scriptsize II} region, $t_{\rm s}=l/v_{\rm s}$, 
where $l$ is the squeezing distance and $v_{\rm s}$ is the squeezing velocity of expanding H\,{\scriptsize II} region traced by H$\alpha$\,RRL.
The squeezing distance of the large semicircle is around 5\,pc (see Fig.\,\ref{fig9}), which is estimated as extending from 
clump C2 to the perpendicular distance of the line connecting clump C1 and C3. The squeezing distance of another small semicircle 
is around 2.4\,pc. The squeezing velocities on the large and small semicircles are equal to the velocity gradient of 
H\,{\scriptsize II} regions as 1.1 and 0.5 km\,s$^{-1}$\,pc$^{-1}$, respectively (see Fig.\,\ref{fig8}).
The two semicircle structures could form at a similar time and forming timescale is estimated around 5\,Myr.

When an H\,{\scriptsize II} region expands at the border of filament G37 (see Fig.\,\ref{fig1}), it can initiate the formation of stars in its vicinity.
H\,{\scriptsize II} region creates variations in density and pressure within the surrounding interstellar medium. 
These variations can cause molecular clouds, which are dense regions of gas, to collapse under their gravity. 
As the clouds collapse, they fragment into smaller clumps, eventually forming protostars. The high-energy radiation 
emitted by the H\,{\scriptsize II} region can also heat up the surrounding gas, further promoting the collapse and 
fragmentation process. Ultimately, this leads to the birth of new stars within the expanding H\,{\scriptsize II} region.
As stated in Sect.\,\ref{Overview}, the filament G37 has been subject to ongoing observations of star formation activities. 
The fragmentation at the filament body of G37 is similar to the shell fragmentation of the Collect and Collapse (CC) model \citep{2011A&A...527A..62B}. Clumps C1, C3, and C4 represent the nearest structures along the filament to the H\,{\scriptsize II} region, which may exhibit an age greater than that of the other clumps and the additional fragmentation observed within the filament (as detailed in Tab.\,\ref{B1}). This proximity suggests a potential association with sequential star formation, as described in the CC model \citep{1994MNRAS.268..291W}.
It is postulated that the expanding H\,{\scriptsize II} region could serve as a significant trigger mechanism for the star formation processes transpiring within the filament G37.
The phenomena observed in G37 are analogous to the CC model of triggered star formation occurring in dense shells surrounding H\,{\scriptsize II} regions \citep{1994MNRAS.268..291W,2008A&A...482..585D,2011A&A...527A..62B}.

\subsection{Impact of the Magnetic Field} 
The magnetic field is a pivotal factor in the physical evolution of molecular clouds (e.g., \citealt{2012ARA&A..50...29C,2019FrASS...6....3H,2021Galax...9...41L,2024ApJ...967...18Z}). 
It may play a significant role in governing the collapse and fragmentation processes within these clouds, thereby impacting the 
formation of stars. Furthermore, the magnetic field influences the distribution and movement of gas and dust within the cloud, 
thereby molding its structure and density. We investigate the role that the magnetic field plays in the formation and evolution 
of filament G37 at the filament scale. As shown in Fig.\,\ref{fig6}, the filament-parallel-alignment magnetic field maintains 
the curved structure of the filament and counteracts the pressure exerted by the expanding H\,{\scriptsize II} region, 
which the clear magnetic field structure can be derived by VGT technique on CO lines with a special velocity range of 
[51,64] km\,s$^{-1}$ in this work (see Sect.\,\ref{VGTm}). In detail, the magnetic field exhibits a distinct curvature and 
aligns parallel to the long axis of filament G37 at the larger semicircle structure of the filament body (see Fig.\,\ref{fig6}), 
which is similar to the curved magnetic field squeezed by surrounding expanding H\,{\scriptsize II} regions in bubble N4 and 
NGC\,6334 molecular complex \citep{2017ApJ...838...80C,2021A&A...647A..78A,2023ApJ...944..139T}.
Due to the resolution of the magnetic field being below to the scale of the smaller semicircle, the magnetic field is barely distorted.
A future high-resolution observation may unveil the detailed magnetic field structure of the smaller semicircle within filament G37.

\begin{figure}
\centering
\includegraphics[width = 9.9cm]{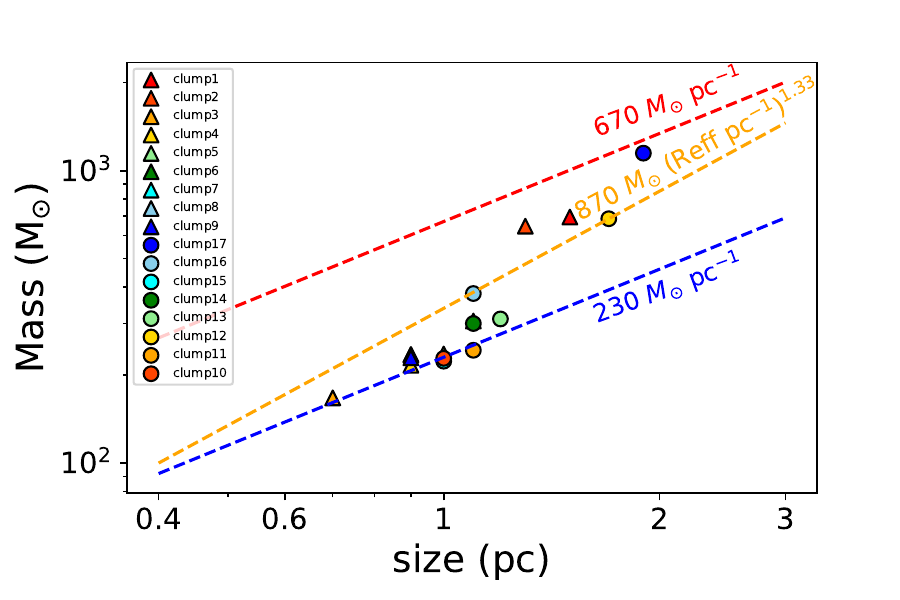}
\caption{The size and mass of clumps are depicted using colorful circles and triangles, as shown in Fig.\,\ref{B1}, on the size-mass map. 
The size refers to the effective diameter of the clumps, which is twice the effective radius. The red and blue dot lines represent the line 
masses of 230 and 670\,M$_\sun$\,pc$^{-1}$ respectively, which serve as representative values for non-star-forming and star-forming 
filaments \citep{2016A&A...591A...5L}. The orange lines illustrate the relationship associated with massive star 
formation \citep{2010ApJ...712.1137K}.}
\label{fig10}
\end{figure}

As delineated in Sect.\,\ref{Material-Flows}, the observation of material flowing in multiple directions along filament G37 was noted.
The curved magnetic field in filament G37 may play a significant role in guiding the material flows within the filament body.
Specifically, the magnetic field aligns parallel to the long axis of the filament G37, which coincides with the direction of material flows.
This parallel alignment ensures that the material remains unaffected by the magnetic field force, as it flows in the same 
direction. Conversely, if the orientation of material flow were perpendicular to both the magnetic field and the filament body, 
the material would experience a magnetic field force opposing its motion. This highlights the ability of the magnetic field to 
influence and guide the material flow within the filament. By guiding the material flow, the magnetic field actively contributes 
to preserving the curved filament structure. 

In contrast to bubble N131 influenced by the H\,{\scriptsize II} region, the molecular cloud is susceptible to fragmentation 
and disruption under the pressure exerted by the expanding H\,{\scriptsize II} region, as demonstrated by \cite{2016A&A...585A.117Z}. 
However, the magnetic field structure of filament G37 exhibits a curved configuration, where the magnetic field effectively 
withstands the pressure exerted by the expanding H\,{\scriptsize II} region on the filament body.
The curved magnetic field structure at the filament scale maintains the filament structure and preserves the structural integrity 
of filament G37.

\begin{figure*}
\centering
\includegraphics[width = 19cm]{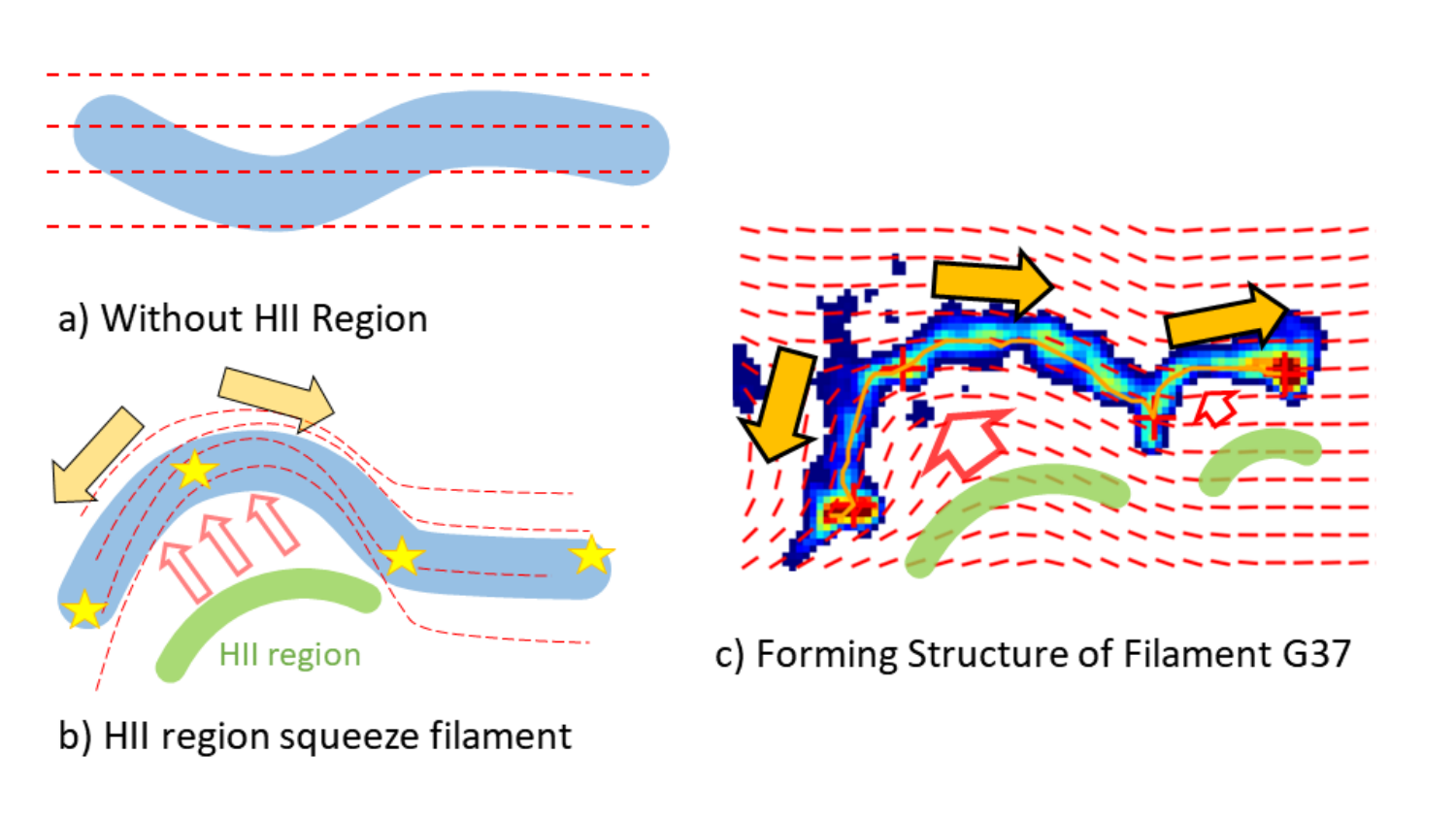}
\caption{Schematic formation diagram proposed for filament G37. The main structure of filament G37 underwent three evolutionary phases, namely: 
a) a filament lacking an H\,{\scriptsize II} region, b) a filament compressed by an expanding H\,{\scriptsize II} region, and 
c) the formation of filament G37. The structure of the filament is depicted by the blue lines, while the $B$-field is represented 
by the red dashed lines. The position of the H\,{\scriptsize II} region is illustrated by the green graphics. Additionally, 
the yellow stars and red crosses indicate the locations where star formation was triggered.}
\label{fig11}
\end{figure*}

We also estimate the mean magnetic field strength and Alfvén Mach number of G37 to be around 23.3 $\mu$G and 2.0, respectively 
(see Sect.\,\ref{apa31} and Table\,\ref{tab1}). The Alfvén Mach number 
($M_{\rm A}$ = $\frac{\sqrt{4\pi \rho}\sigma_v}{B}$ = $\sqrt{\frac{E_{\rm k}}{E_{\rm B}}} \approx 2$) presents that the mean kinetic energy 
density $E_{\rm k}$ in the filament is slightly larger than the local magnetic energy density $E_{\rm B}$.
This suggests that the local turbulent motion related to kinetic energy could affect the average magnetic field structure of filament G37.

\subsection{Star Formation in Filament G37}
As mentioned in Sect.\,\ref{Overview}, the filamentary structure of G37 exhibits bright 8\,$\mu$m emission, especially in four 
dense clumps C1-C4 (see Fig.\,\ref{fig1}), which has the potential 
to initiate star formation. To investigate star formation within filament G37, we have identified 17 clumps within its filamentary body
(see Sect.\,\ref{clumps}). Fig.\,\ref{fig10} illustrates the correlation between the size and mass of these clumps. 
The line mass of the clumps in filament G37 consistently surpasses the lower threshold value of the critical mass for star-forming clouds, 
which is 230\,M$_{\odot}$\,pc$^{-1}$ \citep{2016A&A...591A...5L}. 
This suggests the likelihood of star formation activity within these clumps. 
Moreover, clumps 1, 2, 12, 16, and 17 exhibit a line mass exceeding the threshold value for massive star formation, specifically 870\,M$_{\odot}$\,(R$_{\text{eff}}$\,pc$^{-1.33}$) \citep{2010ApJ...712.1137K}. These dense clumps (1, 2, 12, 16, and 17) could be 
considered as potential candidates for massive star formation. 

Finding massive clumps in filament G37 is necessary, where it may undergo gravitational collapse.
Figs.\,\ref{fig1} and \ref{B1} indicate that the candidates for massive star formation, namely clumps 1, 2, 12, 16, and 17, 
are located within clumps C1, C2, C3, and C4. Furthermore, Fig.\,\ref{fig3} reveals a shift in the direction of material flow 
within filament G37, occurring at clumps C1, C2, C3, and C4. This suggests that the formation of dense clumps 
(C1, C2, C3, and C4) could be attributed to the accretion of material, resulting in material flow in various directions 
within the filamentary structure. Specifically, four dense clumps (1, 2, 12, and 17) situated at the end of filament G37 
exhibit the highest masses among all identified clumps (see Table\,\ref{tab2}). These results further support the notion that star formation 
within filament G37 is characterized by end-dominated collapse.

\subsection{Schematic Formation of Filament G37}
The schematic formation diagram proposed for filament G37 is depicted in Fig.\,\ref{fig11}, highlighting three distinct phases. 
The first phase involves a filament devoid of an H\,{\scriptsize II} region, followed by a phase where the filament experiences 
compression due to the expansion of an H\,{\scriptsize II} region. 
Finally, the formation of filament G37 occurs as a result of the magnetic field resistance against the compression exerted by the H\,{\scriptsize II} regions. 
During the initial phase without an H\,{\scriptsize II} region, the giant filament maintains a close alignment with the Galactic plane.
Previous observations of giant filaments, such as G11, G29, G51, and the Radcliffe Wave, reveal an "S"-shaped morphology within the Milky Way \citep{2013A&A...559A..34L,2015MNRAS.450.4043W,2022MNRAS.517L.102L,2024Natur.628...62K}. Therefore, we hypothesize that the G37 filament initially exhibits a similar "S"-shaped structure, independent of the influence of the expanding H\,{\scriptsize II} region (see Fig. 8).
The local magnetic field within this filament is observed to be parallel to its body, as documented by \cite{2018ApJ...864..153Z}, 
\cite{2021A&A...651L...4S}, and \cite{2024ApJ...961..124Z}.
Due to the insufficient evidence available for comparing the ages of filaments and the expansion timescales of H\,{\scriptsize II} regions, it is challenging to infer the sequential relationship between filament formation and H\,{\scriptsize II} region expansion. In this context, filaments devoid of associated H\,{\scriptsize II} regions serve as a control group, allowing for the investigation of the impact of H\,{\scriptsize II} regions on filamentary structures.

During the evolutionary phase involving an H\,{\scriptsize II} region, the expansion of these regions exerts significant pressure, 
resulting in the distortion of the filament body, as visually demonstrated in Fig.\,\ref{fig11}. The local magnetic field is also 
subject to distortion due to the pressure exerted by the expanding H\,{\scriptsize II} region, leading to changes in alignment along 
the filament body and the formation of curved structures \citep{2017ApJ...838...80C,2021A&A...647A..78A}. The compression caused by 
the expanding H\,{\scriptsize II} regions induces the accumulation of gas into dense clumps, which subsequently triggers the process 
of star formation. As dense clumps accrete material, there are flows of material towards these clumps, resulting in multi-directional 
material flows along the filament body.

As the evolution of the filament G37 progresses, the magnetic field within the filament plays a crucial role in counteracting the pressure exerted by the expanding H\,{\scriptsize II} region. 
This interaction continues until a dynamic equilibrium is achieved within the system. Throughout this process, the curved structure of filament G37 undergoes formation and reshaping. The presence 
of the nearby expanding H\,{\scriptsize II} region exerts compression on the filament, resulting in a curved filament body and 
magnetic field structure. This phenomenon triggers the formation of stars and gives rise to the unique occurrence of multi-directional 
material transportation within filament G37 during its evolutionary stages. The curved magnetic field provides support against the 
compression exerted by the H\,{\scriptsize II} region, thereby maintaining the multiple directions of material flow along the filament 
body caused by clump accretion, as well as preserving the curved structure of the filament. This physical process, characterized by the interplay between the magnetic field, H\,{\scriptsize II} region, and material flows, can persist for a duration of up to 5\,Myr.

\section{Summary}
\label{sect5}
We conducted an investigation into the dynamics and magnetic field characteristics of filament G37, while also delving into the physical
mechanisms underlying its formation. The main results are the following:

\begin{enumerate}
\item
The filament G37 exhibits a distinct influence on the motion of material, directing it towards four dense clumps situated at specific locations:
C1 and C4, positioned at the ends of the filament, C2, located at the apex of its large semicircular structure, and C3, situated between
the two semicircular components. These four clumps present themselves as potential regions for the formation of massive stars.

\item
By employing the $^{13}$CO\,(3-2) spectral line, we have estimated the velocity gradients of the material
flowing from clumps C2 to C1, C2 to C3, and C3 to C4 to be approximately 0.46, 0.40, and 0.49\,km\,s$^{-1}$\,pc$^{-1}$, respectively. 
The observed multi-directional flows of material within these clumps indicate that they may be attributed to the accretion processes 
involving these substantial clumps. The filament G37 provides a potential candidate for end-dominated collapse.

\item
The velocity gradients of the two expanding H\,{\scriptsize II} regions, which are associated with the large and small semicircles of filament G37, 
were determined to be $\sim$1.10 and 0.54\,km\,s$^{-1}$\,pc$^{-1}$, respectively. Through the analysis of tracers such as $^{12}$CO, $^{13}$CO, 
and H$\alpha$\,RRL, it was observed that the expanded H\,{\scriptsize II} region, as traced by H$\alpha$\,RRL, induces compression of the 
filament gas in the direction of the blue shift. The formation of two semicircular structures within filament G37 is likely attributable to the
nearby expanding H\,{\scriptsize II} regions. This compression effect could potentially serve as a triggering mechanism for star
formation within the filament.

\item
The measured magnetic field obtained from VGT observations revealed a distinctive curved structure, aligning parallel to the elongated axis
of filament G37. Despite the compressive effects exerted by the adjacent H\,{\scriptsize II} region, the magnetic field within the filament 
G37 plays a crucial role in preserving its current morphology and directing the flow of material within the filamentary structure. 
The estimated strength of the magnetic field yielded a value of $\sim$23\,$\mu$G.

\item
The combined effects of H\,{\scriptsize II} region expansion and the curved magnetic field contribute 
to the unique morphology of G37, resembling a "caterpillar" shape. The timescale for these physical processes within G37 is estimated to 
be $\sim$5\,Myr.
\end{enumerate}

\begin{acknowledgements}
The authors thank the anonymous referee for helpful comments.
We thank Dr. Yue Hu and Prof. Alex Lazarian for the VGT code and helpful comments. 
This work acknowledges the support of the National Key R\&D Program of China under grant Nos.\,2023YFA1608002, 2023YFA1608204, and 2022YFA1603100, the Chinese Academy of
Sciences (CAS) “Light of West China” Program under grant Nos.\,xbzg-zdsys-202212, the Tianshan Talent Program of Xinjiang 
Uygur Autonomous Region under grant No.\,2022TSYCLJ0005, the Natural Science Foundation of Xinjiang Uygur Autonomous Region under grant No.\,2022D01E06, the Xinjiang Key Laboratory of Radio Astrophysics under grant No.\,2023D04033, the National Natural Science Foundation 
of China under grant Nos.\,12173075, 12425304, and U1731237, and the Youth Innovation Promotion Association CAS. This research has used NASA's Astrophysical Data System (ADS).
\end{acknowledgements}

\bibliographystyle{aa}
\bibliography{reference}

\newpage

\appendix


\section{Averaged Spectral Lines of Filament G37}
\begin{figure}[h]
\centering
\includegraphics[width = 9cm]{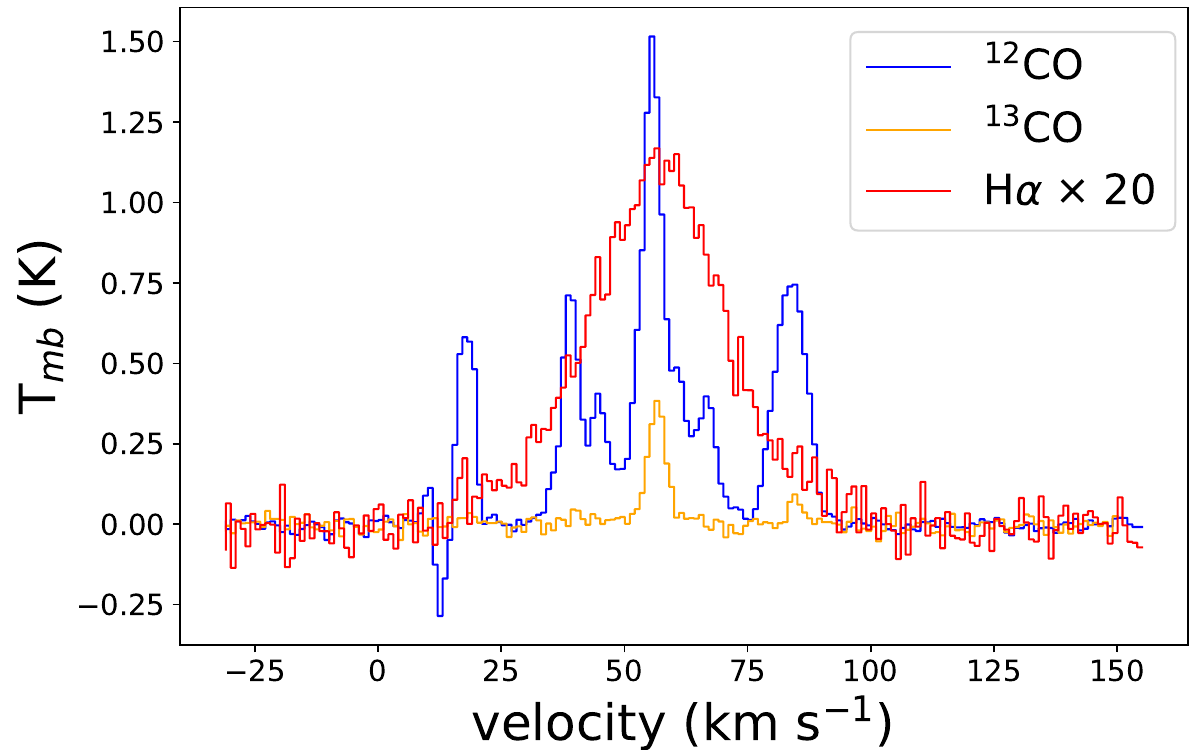}
\caption{Three mean spectral lines of $^{12}$CO\,(3-2), $^{13}$CO\,(3-2), and H$\alpha$\,RRL are observed in filament G37. The $^{12}$CO\,(3-2) spectrum 
is represented by the blue line, while the $^{13}$CO\,(3-2) spectrum is depicted by the orange line. Additionally, the H$\alpha$\,RRL spectrum, 
with its intensity multiplied by 20 times, is shown by the red line.}
\label{fig7}
\end{figure}

\section{Distributions of Central Velocities and Velocity Dispersions of CO}
\begin{figure*}[h]
\centering
\includegraphics[width = 18cm]{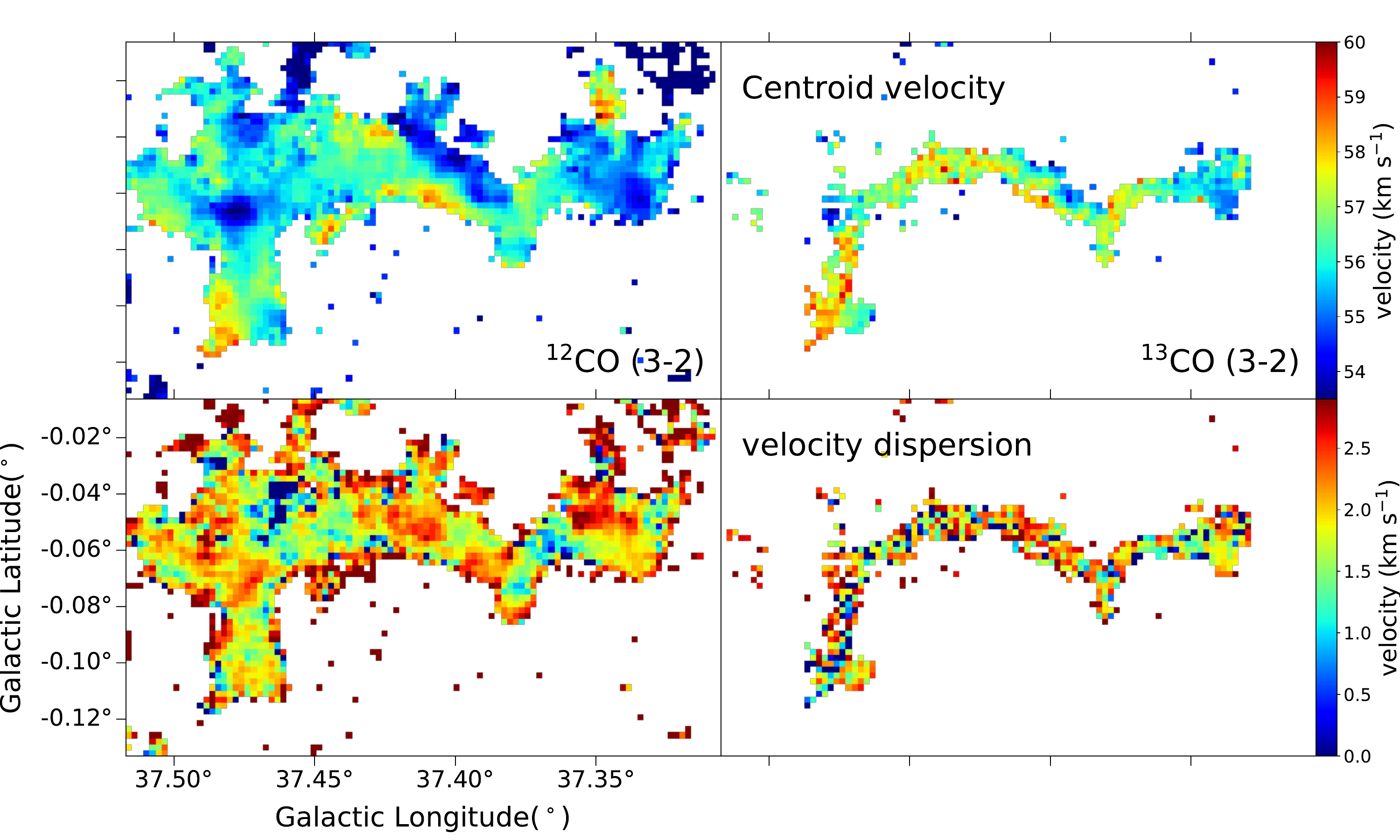}
\caption{The distributions of central velocities (\emph{top panels}) and velocity dispersions (\emph{bottom panels}) derived from $^{12}$CO
and $^{13}$CO\,(3-2) spectra.}
\label{fig4}
\end{figure*}

\section{Magnetic Field Measured with Velocity Gradient Technique}
\label{VGTm}
\subsection{Velocity Gradient Technique}
\begin{figure*}[t]
    \centering
    \includegraphics[width = 20cm]{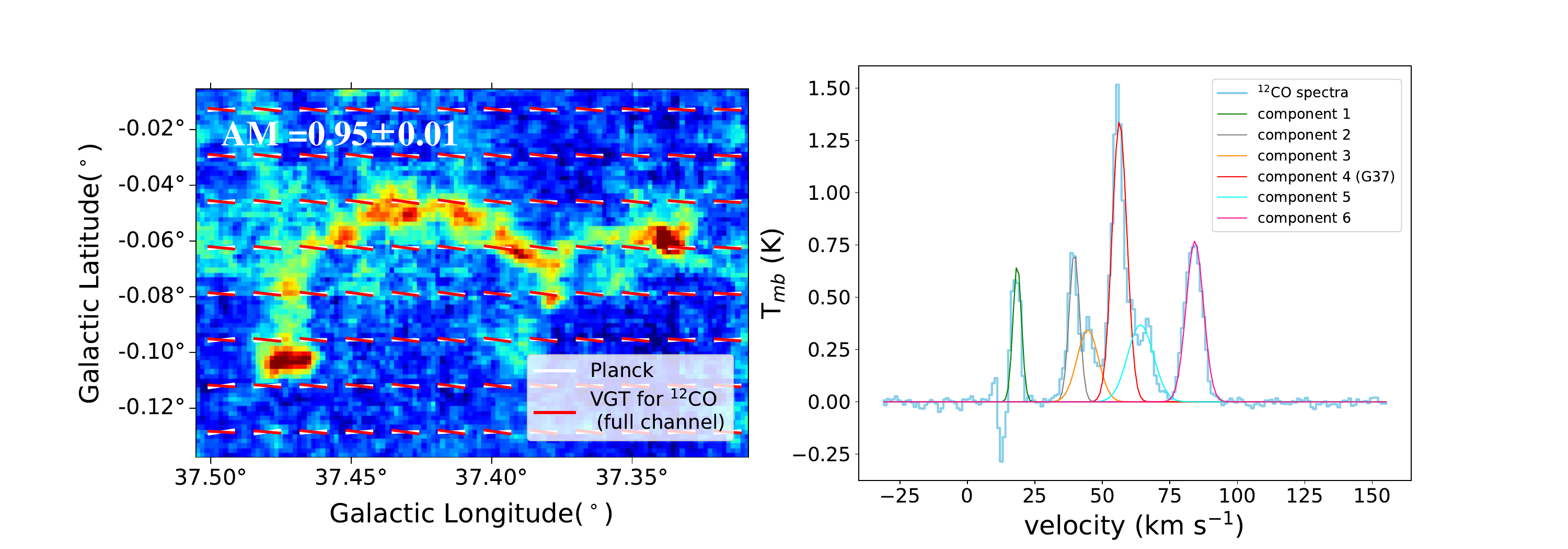}
    \caption{Comparison of magnetic field measured with VGT for $^{12}$CO\,(3--2) and $Planck$ dust polarization.
    The comparison of magnetic fields measured with VGT for $^{12}$CO\,(3--2) and $Planck$ dust polarization is shown in the left panel. 
    The orientation of the magnetic field measured using VGT for $^{12}$CO\,(3--2) spectra and dust polarization from $Planck$ is 
    displayed by the red and white vectors, respectively. The integrated intensity of  $^{12}$CO\,(3--2) spectral line from 
    -31 to 155\,km\,s$^{-1}$ is shown in the background. The right panel shows the spectral line of $^{12}$CO\,(3--2) across the full 
    velocity channel, with the six different velocity components represented by different color lines.
    }
    \label{fig5}
\end{figure*}

As depicted in Fig.\,\ref{fig5}, six distinct velocity components were identified across various velocity ranges, with Component 4 constituting a segment of filament G37. This observation implies the presence of multiple objects within the line of sight (LOS) direction traversing the multi-spiral arms. The continuum observation is significantly influenced by a substantial number of foreground and background objects originating from different spiral arms. Consequently, measuring the magnetic field of filament G37 using dust-polarized continuum becomes challenging due to the inclusion of these foreground and background objects. The removal of these foreground effects proves to be particularly difficult.

The velocity gradient technique (VGT; \citealt{2017ApJ...835...41G,2018ApJ...853...96L,2018MNRAS.480.1333H}) offers a method to measure magnetic fields with multiple velocity components. By utilizing the position-position-velocity (PPV) cubes derived from spectroscopic data, these velocity components can be distinctly separated, allowing for accurate measurements of their respective magnetic fields. Employing the $^{12}$CO\,(3-2) spectral line in conjunction with the VGT technique facilitates the acquisition of a pristine magnetic field structure for filament G37, unaffected by foreground and background interferences.

In this study, the VGT technique is employed for analyzing the magnetic field. This method is predicated on the anisotropy of magneto-hydrodynamic turbulence \citep{1995ApJ...438..763G} and fast turbulent reconnection theories \citep{1999ApJ...517..700L}. In order to extract the velocity information from PPV cubes, thin velocity channels denoted as Ch(x,y) were utilized,
\begin{equation}
    \begin{aligned}
        \rm\bigtriangledown_x Ch_i(x,y) = Ch_i(x,y) - Ch_i(x-1,y) \,,
    \end{aligned}
\end{equation}
\begin{equation}
    \begin{aligned}
        \rm\bigtriangledown_y Ch_i(x,y) = Ch_i(x,y) -Ch_i(x,y-1)\,, 
    \end{aligned}
\end{equation}
\begin{equation}
    \begin{aligned}
        \rm\psi^i_g = tan^{-1} (\frac{\bigtriangledown_y Ch_i(x,y)}{\bigtriangledown_x Ch_i(x,y)} \,,
    \end{aligned}
\end{equation}
where $\bigtriangledown_x$Ch$_i$(x,y) and $\bigtriangledown_y$Ch$_i$(x,y) are the x and y components of the gradient, respectively. 
This procedure is executed for pixels exhibiting a spectral line emission with a signal-to-noise ratio (SNR) exceeding 3.

The orientation of the magnetic field is perpendicular to the velocity gradient, and these velocity gradients must exhibit statistical significance. A technique known as sub-block averaging \citep{2017ApJ...837L..24Y} has been employed to extract velocity gradients from raw gradients within a designated sub-block of interest. Subsequently, a histogram corresponding to the raw velocity gradient orientations, denoted as $\psi^i_{\rm g}$, is plotted. The size of this sub-block is fixed at 20$\times$20 pixels, which also dictates the resolution of the final magnetic field. Through the application of sub-block averaging, eigen-gradient maps $\psi^i_{g_s}(x,y)$ are obtained and used to compute the pseudo-Stokes-parameters $Q_{\rm g}$ and $U_{\rm g}$. Consequently, the pseudo-Stokes-parameters $Q_{\rm g}$ and $U_{\rm g}$ for the inferred magnetic field are constructed based on these parameters by:

\begin{equation}
    \begin{aligned}
        \rm Q_g(x,y) = \sum\limits_{i=1}^{n_v} I_i(x,y)cos(2\psi^i_{g_s}(x,y))  ,
    \end{aligned}
\end{equation}
\begin{equation}
    \begin{aligned}
        \rm U_g(x,y) = \sum\limits_{i=1}^{n_v} I_i(x,y)sin(2\psi^i_{g_s}(x,y)) ,
    \end{aligned}
\end{equation}
\begin{equation}
    \begin{aligned}
        \rm\psi_g = \frac{1}{2}tan^{-1}\frac{U_g}{Q_g} ,
    \end{aligned}
\end{equation}
where $\psi_{\rm g}$ is the pseudo polarization angle. The pseudo polarization angle  is perpendicular to the POS orientation angle 
of the magnetic field: $\psi_B = \psi_g + \pi /2$.



To evaluate the precision of the VGT in tracing magnetic fields, we utilized the PPV cubes from the full velocity channels of the $^{12}$CO\,(3-2) spectral line (see Fig.\,\ref{fig5}). This was done to compare the B-field derived from $Planck$ 353\,GHz dust polarization with that obtained using VGT. The $^{12}$CO\,(3-2) emission was chosen for this application due to its superior SNRs compared to the $^{13}$CO\,(3-2) emission. Notably, the $^{12}$CO\,(3-2) emission spans the entire velocity range, encompassing all components along the line of sight. When applied to the VGT method using this $^{12}$CO\,(3-2) emission, the resultant magnetic field encompasses both foreground and background components. Its origin aligns closely with that of the polarized dust continuum, such as the $Planck$ 353\,GHz dust polarization.
In order to juxtapose the magnetic field as measured by VGT and Planck continuum, we established a sub-block size of 50$\times$50 pixels. This was done to align with the same beam as that utilized by $Planck$ ($\sim$5$'$). The alignment between the B-field orientations, as detected through polarization $\theta_B$ and VGT $\Phi_B$, is characterized by the Alignment Measure (AM, \citealt{2017ApJ...835...41G}): $\rm AM = 2(\left \langle cos^2(\theta_B - \Phi_B) \right \rangle - \frac{1}{2})$. The AM values span a range from -1 to 1. An AM value nearing 1 suggests that $\phi_B$ is parallel to $\psi_g$, whereas an AM value approaching -1 implies that $\phi_B$ is perpendicular to $\psi_g$. The uncertainty associated with the AM value, denoted as $\sigma_{AM}$, can be determined by dividing the standard deviation by the square root of the sample size. This calculation is illustrated in Fig.\,\ref{fig5}.

As depicted in Fig.\,\ref{fig5}, the AM value reaches up to 0.95\,$\pm$\,0.01. This indicates that the magnetic field, as measured by VGT in full velocity channels of the $^{12}$CO\,(3-2), closely aligns with that of $Planck$, exhibiting a mean offset angle of less than 10$^\circ$ between the two types of magnetic field orientation angles. The ability of VGT to trace the magnetic field in filament G37 region is consistent with that of $Planck$, demonstrating high accuracy. When compared to dust polarization, VGT effectively distinguishes between the foreground and background using a single velocity component \citep{2018ApJ...853...96L,2019NatAs...3..776H,2022MNRAS.510.4952L,2022MNRAS.513.3493H,2022ApJ...934...45Z}.

\subsection{Magnetic Field Strength}
\label{apa31}
\citealt{2020arXiv200207996L} introduce a novel method, termed the MM2 technique, for estimating the magnetic field strength $\it{B}_{\rm pos}$ in the plane-of-sky (POS). This technique is specifically applied to both the sonic Mach number and the Alfv\'en Mach number. The sonic Mach number can be determined using spectral lines and dust temperature measurements:
\begin{equation}
    \begin{aligned}
        M_{\rm s}\,=\,\sqrt{3}\sigma_{{\rm v,1D}}/c_{\rm s},
    \end{aligned}
\end{equation}
where $\sigma_{\rm v,1D}$ is the velocity dispersion obtained from $^{13}$CO\,(3--2), $c_{\rm s}$ is the sonic speed ($c_{\rm s}$\,=\,$\sqrt{k_{\rm B}T/\mu_{\rm p}m_{\rm H}}$, where $T$ is the dust temperature, $k_{\rm B}$ is the Boltzmann constant, $\mu$\,=\,2.37 is the mean molecular weight). 
Based on the dispersion relation in the direction of the velocity gradient shows a power-law arrangement with $M_{\rm A}$ 
\citep{2020arXiv200207996L,2021ApJ...912....2H}, Alfv\'en Mach number $M_{\rm A}$ could be measured by "top-to-bottom" ratio 
of the distribution of the channel velocity gradients (VChGs): 
\begin{equation}
	\label{eq.tb}
	\begin{aligned}
		M_{\rm A}&\approx1.6 (T_{\rm v}/B_{\rm v})^{\frac{1}{-0.60\pm0.13}}, \,\,\,\,\,for\,\,\,M_{\rm A}\le1 \,, \\
		M_{\rm A}&\approx7.0 (T_{\rm v}/B_{\rm v})^{\frac{1}{-0.21\pm0.02}}, \,\,\,\,\,for\,\,\,M_{\rm A}\textgreater 1 \,, \\
	\end{aligned}
\end{equation}
where $T_{\rm v}$ represents the maximum value of the fitted histogram in the velocity gradient directions, while $B_{\rm v}$ is the minimum value of that. Using the MM2 technique, magnetic field strength $\it{B}_{\rm pos}$ can be calculated by:
\begin{equation}
    B=\Omega c_{\rm s}\sqrt{4\pi\rho_0}\,M_{\rm s}M_{\rm A}^{-1},
\end{equation}
where the $\Omega$ is a geometrical factor ($\Omega$ = 1), $\rho_0$ is the volume density estimated by long uniform cylinder model \citep{2000MNRAS.311...85F,2017ApJ...846..122P}. 
$M_{\rm A}$ measured by VChGs is 1.97. The mean value of $\it{B}_{\rm pos}$ in filament G37 is estimated as 23.3\,$\mu$G. 
The detail of other physical parameters of filament G37 is shown in Table\,\ref{tab1}.


\section{Maps of $^{12}$CO velocity components}
\label{ApE}


\begin{figure*}
    \centering
    \includegraphics[width=18.5cm]{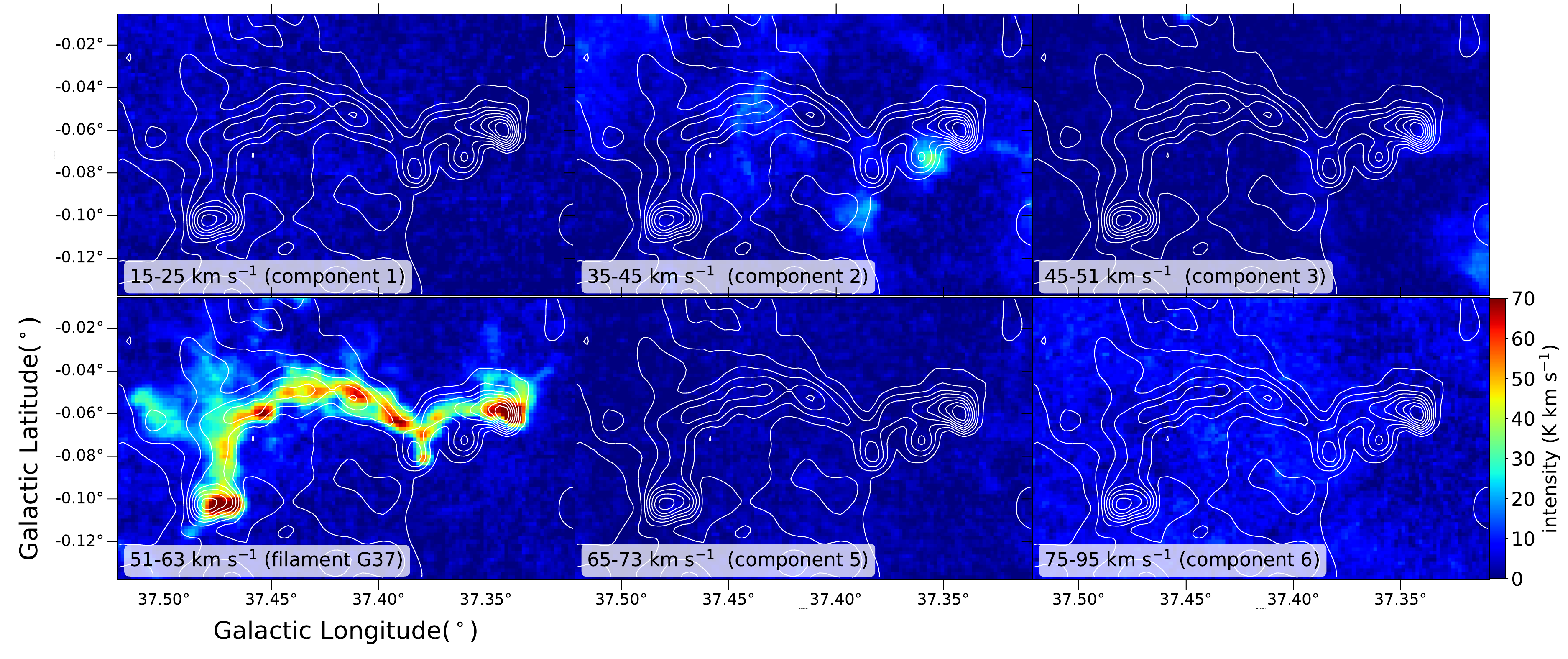}
    \caption{Velocity-integrated intensity maps of $^{12}$CO for each velocity component (see Fig.\,\ref{fig5}) are presented in the background.
    The white contours show the H$_2$ column density, ranging from 10$^{22}$ to 10$^{22.4}$\,cm$^{-2}$ with a step of 10$^{0.05}$\,cm$^{-2}$ \citep{2018ApJ...864..153Z}.}
    \label{figE}
\end{figure*}

\section{Clumps}
\label{ApD}

\begin{figure*}[h]
    \centering
    \includegraphics[width = 18.5cm]{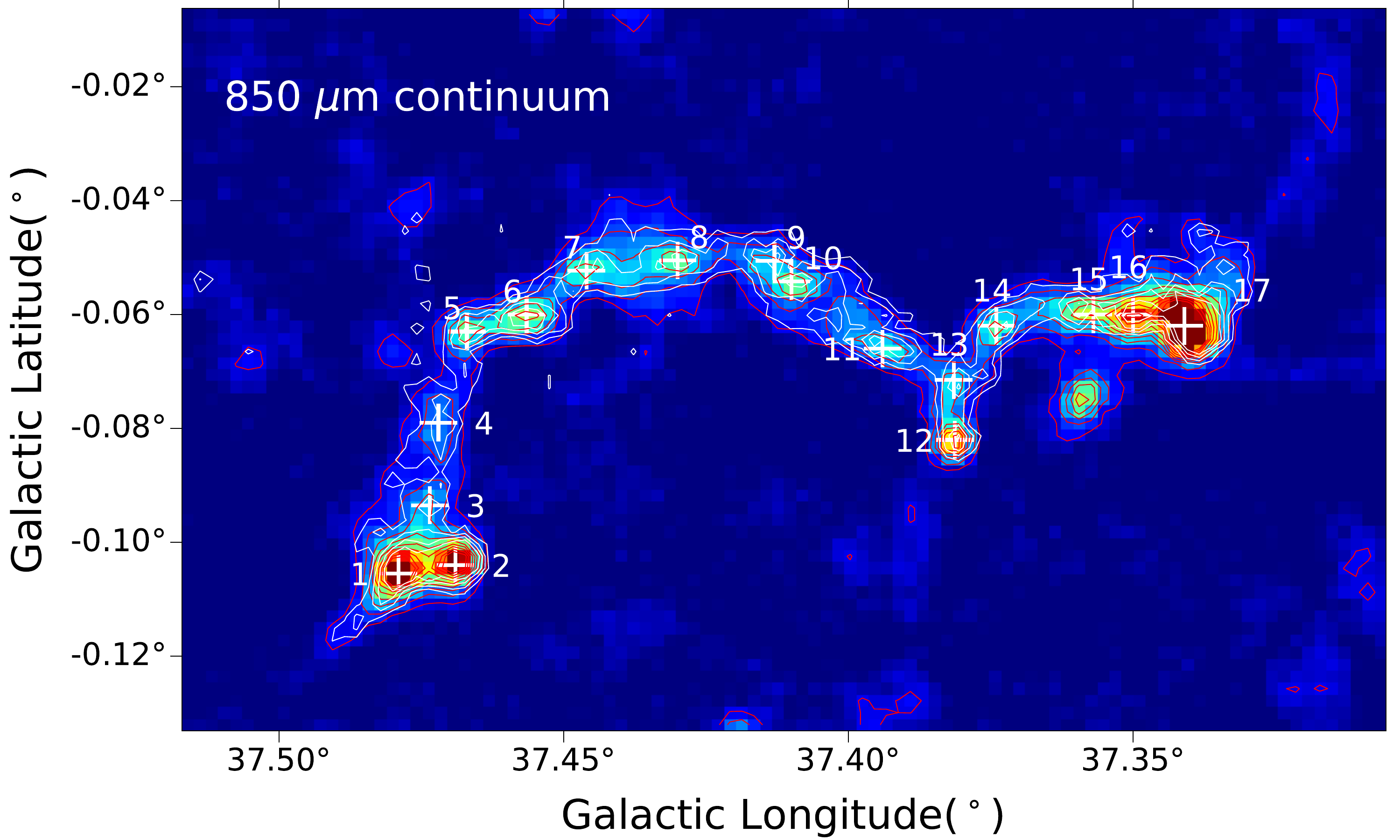}
    \caption{Identified clumps marked with white crosses in filament G37. Red contours represent the shape of 850\,$\mu$m emission, which ranges from 50 to 400\,mJy\,beam$^{-1}$ with steps of 50\,mJy\,beam$^{-1}$. The white contours correspond to the $^{13}$CO\,(3-2) emission, consistent with Fig.\,\ref{fig1}.}
    \label{B1}
\end{figure*}

A total of 17 clumps are identified in filament G37 using the 850\,$\mu$m continuum, as shown in Fig.\,\ref{B1} and Table\,\ref{tab2}.
We utilized the $^{13}$CO\,(3--2) emission of G37 to disentangle clumps, ensuring that it aligned with the shape of the 850\,$\mu$m emission from cold dust simultaneously. Based on the emission contours from the 850\,$\mu$m continuum, we identified the positions and structures of the clumps. The mass of each clump was estimated using the H$_2$ column density in conjunction with the clump structure derived from the $^{13}$CO\,(3--2) observations.

\begin{table}[h]
    \centering
    \caption{Physical Parameters of Clumps}
    \begin{tabular}{c c c c c c }
    \hline
    \hline
    Clump & $l$ & $b$ & $N$(H$_2$) & Mass & Size \\
         & $^\circ$ & $^\circ$ & 10$^{22}$ cm$^{-2}$ & M$_\sun$ & pc \\
    \hline
    1 (C1)  & 37.479 & -00.106 & 1.7 & 694  & 1.5 \\
    2 (C1)  & 37.469 & -00.104 & 1.6 & 646  & 1.3  \\
    3       & 37.474 & -00.093 & 1.6 & 167  & 0.7  \\
    4       & 37.472 & -00.079 & 1.4 & 216  & 0.9\\
    5       & 37.467 & -00.063 & 1.5 & 236  & 1.0\\
    6 (C2)  & 37.457 & -00.058 & 1.5 & 306  & 1.1 \\
    7       & 37.446 & -00.052 & 1.5 & 235  & 0.9\\
    8       & 37.431 & -00.051 & 1.5 & 234  & 0.9 \\
    9       & 37.413 & -00.051 & 1.5 & 229  & 0.9\\
    10      & 37.411 & -00.054 & 1.5 & 228  & 1.0\\
    11      & 37.394 & -00.066 & 1.2 & 243  & 1.1\\
    12 (C3) & 37.381 & -00.082 & 1.3 & 685  & 1.7\\
    13      & 37.382 & -00.071 & 1.3 & 311  & 1.2\\
    14      & 37.374 & -00.062 & 1.3 & 300  & 1.1\\
    15      & 37.357 & -00.061 & 1.4 & 223  & 1.0\\
    16 (C4) & 37.351 & -00.062 & 1.6 & 380  & 1.1\\
    17 (C4) & 37.341 & -00.062 & 1.8 & 1148 & 1.9\\
    \hline
    \hline 
    \end{tabular}
    \label{tab2}
\end{table}

\end{document}